\documentclass[fleqn,usenatbib]{mnras}

\usepackage{newtxtext,newtxmath}
\usepackage[T1]{fontenc}

\DeclareRobustCommand{\VAN}[3]{#2}
\let\VANthebibliography\thebibliography
\def\thebibliography{\DeclareRobustCommand{\VAN}[3]{##3}\VANthebibliography}

\usepackage{graphicx}
\usepackage{amsmath}
\usepackage{comment}

\title[Cosmological baryon spread in CAMELS]{Cosmological baryon spread and impact on matter clustering in CAMELS}

\author[Gebhardt et al.]{Matthew Gebhardt,$^{1}$\thanks{E-mail: matthew.gebhardt@uconn.edu}
Daniel Anglés-Alcázar$^{1,2}$, 
Josh Borrow$^{3,4}$, 
Shy Genel$^{2,5}$, 
\newauthor 
Francisco Villaescusa-Navarro$^{2,6}$, 
Yueying Ni$^{7,8}$,
Christopher Lovell$^{9, 10}$,
Daisuke Nagai$^{11}$, 
\newauthor
Romeel Davé$^{12,13,14}$, 
Federico Marinacci$^{15}$, 
Mark Vogelsberger$^{3,16}$, 
Lars Hernquist$^{7}$
\\
$^{1}$Department of Physics, University of Connecticut, 196 Auditorium Road, U-3046, Storrs, CT 06269-3046, USA\\
$^{2}$Center for Computational Astrophysics, Flatiron Institute, 162 Fifth Avenue, New York, NY 10010, USA\\
$^{3}$Department of Physics and Kavli Institute for Astrophysics and Space Research, Massachusetts Institute of Technology, Cambridge, MA 02139, USA\\
$^{4}$Institute for Computational Cosmology, Department of Physics, Durham
University, South Road, Durham DH1 3LE, UK\\
$^{5}$Columbia Astrophysics Laboratory, Columbia University, 550 West 120th Street, New York, NY 10027, USA\\
$^{6}$Department of Astrophysical Sciences, Princeton University, 4 Ivy Lane, Princeton, NJ 08544 USA\\
$^{7}$Harvard-Smithsonian Center for Astrophysics, 60 Garden Street, Cambridge, MA 02138, USA\\
$^{8}$McWilliams Center for Cosmology, Department of Physics, Carnegie Mellon University, Pittsburgh, PA 15213, USA\\
$^{9}$Institute of Cosmology and Gravitation, University of Portsmouth, Burnaby Road, Portsmouth, PO1 3FX, UK\\
$^{10}$Centre for Astrophysics Research, School of Physics, Engineering \& Computer Science, University of Hertfordshire, Hatfield AL10 9AB, UK\\
$^{11}$Department of Physics, Yale University, New Haven, CT 06520, USA\\
$^{12}$Institute for Astronomy, University of Edinburgh, Royal Observatory, Edinburgh EH9 3HJ, UK\\
$^{13}$Department of Physics \& Astronomy, University of the Western Cape, Cape Town 7535, South Africa\\
$^{14}$South African Astronomical Observatories, Observatory, Cape Town 7925, South Africa\\
$^{15}$Dipartimento di Fisica e Astronomia ‘Augusto Righi’, Universit`a di Bologna, via Gobetti 93/2, 40129, Bologna, Italy\\
$^{16}$The NSF AI Institute for Artificial Intelligence and Fundamental Interactions, Massachusetts Institute of Technology, Cambridge MA\\
}

\date{Accepted XXX. Received YYY; in original form ZZZ}
\pubyear{2023}

\begin{document}
\label{firstpage}
\pagerange{\pageref{firstpage}--\pageref{lastpage}}
\maketitle

\begin{abstract}
We quantify the cosmological spread of baryons relative to their initial neighboring dark matter distribution using thousands of state-of-the-art simulations from the Cosmology and Astrophysics with MachinE Learning Simulations (CAMELS) project.
We show that dark matter particles spread relative to their initial neighboring distribution owing to chaotic gravitational dynamics on spatial scales comparable to their host dark matter halo.
In contrast, gas in hydrodynamic simulations spreads much further from the initial neighboring dark matter owing to feedback from supernovae (SNe) and Active Galactic Nuclei (AGN). We show that large-scale baryon spread is very sensitive to model implementation details, with the fiducial \textsc{SIMBA} model spreading $\sim$40\% of baryons $>$1\,Mpc away compared to $\sim$10\% for the IllustrisTNG and \textsc{ASTRID} models. 
Increasing the efficiency of AGN-driven outflows greatly increases baryon spread while increasing the strength of SNe-driven winds can decrease spreading due to non-linear coupling of stellar and AGN feedback. 
We compare total matter power spectra between hydrodynamic and paired $N$-body simulations and demonstrate that the baryonic spread metric broadly captures the global impact of feedback on matter clustering over variations of cosmological and astrophysical parameters, initial conditions, and galaxy formation models. Using symbolic regression, we find a function that reproduces the suppression of power by feedback as a function of wave number ($k$) and baryonic spread up to $k \sim 10\,h$\,Mpc$^{-1}$ while highlighting the challenge of developing models robust to variations in galaxy formation physics implementation.
\end{abstract}




\section{Introduction}
Investigating the distribution of matter in the Universe reveals many clues about its origin, content, and fate. Cosmological parameters such as the density of matter ($\Omega_{\rm{m}}$) and the present-day linear amplitude of matter fluctuations ($\sigma_{8}$) can be constrained by comparing theoretical predictions to observations from the cosmic microwave background \citep{planck_2020}, galaxy clustering \citep{Cole_2005_galclustcosmo, Eisenstein_2005_galclustcosmo}, and weak lensing surveys \citep{Huang_2021, Hadzhiyska_2021_WL}. In this new age of precision cosmology, simulations have become extremely valuable in the pursuit of tighter constraints on cosmological parameters by comparing their outputs to these surveys. As the next generation of surveys (e.g.  CMB-S4\footnote{\url{https://cmb-s4.org}}, DESI\footnote{\url{https://www.desi.lbl.gov}}, eROSITA\footnote{\url{https://www.mpe.mpg.de/eROSITA}}, Euclid\footnote{\url{https://www.euclid-ec.org}}, and Rubin Observatory\footnote{\url{https://www.lsst.org/}}) provide greater statistical power via larger volumes and greater sensitivity, cosmological simulations must follow suit. As their resolution increases, however, simulations must model smaller scales at which matter clustering can no longer be explained purely by gravitational dynamics. At such scales, processes such as radiative cooling, galactic winds driven by supernovae (SNe), and active galactic nuclei (AGN) feedback play an important role in the evolution of galaxies and directly redistribute baryonic matter over a range of scales \citep{DAA_2017, Borrow_2020}, which can provide an important source of contamination when extracting information from cosmological surveys \citep{van_Daalen_2011, Chisari_2019, Schaye_2023_flamingo}. Unfortunately, many key physical mechanisms in galaxy formation are still not well understood, and so it is a challenge to decouple astrophysical processes from the intrinsic effects of fundamental cosmological parameters on the matter distribution. The uncertainties and computational costs of these baryonic processes relegate their implementation in large-volume hydrodynamic simulations to extensively-tuned free parameters in subgrid models \citep{Somerville_2015}. To extract the maximum amount of cosmological information from future surveys, the effects and uncertainties of these processes must be well accounted for. 

Dark matter only (``$N$-body'') simulations have seen great successes in reproducing the over-arching large scale structure of the Universe and achieving the large volumes (at sufficient resolution) required for comparisons to cosmological surveys \citep{Springel_2005c_nbody, Klypin_2011_nbody, Angulo_2012_nbody}. However, while the dark matter component is responsible for the majority of the gravitational potential to form large structures, baryonic matter is subject to various astrophysical processes and, as a result, does not simply follow the dark matter \citep{Naab_2017, Vogelsberger_2019_sims}. There have been a wide range of efforts to create models that approximate the effects of baryons in such simulations. Empirical models \citep[e.g.][]{Berlind_2002_HOD, Bosch_2007_CLF, Behroozi_2010_SHAM} are computationally efficient and map observable properties of baryons to dark matter haloes without any explicit modeling of baryonic processes. Semi-analytical models (SAMs) are a more physically-motivated approximation method \citep[e.g.][]{Kauffmann_1993_SAM, Somerville_1999_SAM, Croton_2006_SAM, Guo_2011_SAM} that predicts galaxy properties given simulated dark matter halo merger trees by solving bulk equations to track quantities such as gas accretion onto haloes, star formation rates, or gas ejected from galaxies \citep{Baugh_2006_SAMs, Somerville_2015}, but still do not predict the total matter distribution in and around galaxies. Cosmological hydrodynamic simulations \citep[e.g.][]{Hirschmann_2014_magneticum, Schaye_2015_EAGLE, mufasa_Dave_2016, SIMBA_Dave_2019, weinberger_2017_TNGpaper} are the most direct way of modeling the impact of baryonic physics on galaxy evolution and the total matter distribution, but suffer from uncertainties in baryonic physics models. 

The predicted abundance, clustering, and concentration of dark matter haloes differs between $N$-body and hydrodynamic simulations \citep{Cui_2014, Cui_2016_baryons, Lu_2021_baryons, Sorini_2021_baryons, Beltz-Mohrmann_2021_baryons}, and connecting these can approximate the predictive power of hydrodynamic simulations. Two methods for such an approximation are ``halo models,'' which alter the radial density profiles of haloes in $N$-body simulations when calculating the total matter power spectrum to match that of hydrodynamic simulations \citep[e.g.][]{Seljak_2000_halo_model, Sembolini_2013_halo_model, Mead_2015_halo_model} and ``baryonification'' methods, which go a step further to actually alter the 3D distribution particles to match the halo density profiles found in hydrodynamic simulations \citep[e.g.][]{Schneider_2015_baryonification, Schneider_2019_baryonification, Weiss_2019_baryonification}. Halo models are also used in modeling \citep[e.g.][]{Shaw_2010_SZmodeling, Osato_2023_SZmodeling} and interpreting SZ surveys \citep[e.g.][]{Reichardt_2012_SZsurvey, Osato_2018_SZanalysis, Osato_2020_SZanalysis}. However, additional cluster astrophysics, such as the feedback and baryonic effects, must be understood better to realize the statistical power of upcoming SZ surveys \citep{Chisari_2019}.

In practice, to compare theory to observations, one typically computes a ``summary statistic,'' such as the matter power spectrum \citep{Philcox_2020, Ivanov_2020, d_Amico_2020}, which describes how matter is clustering at different spatial scales. Relative to $N$-body simulations, matter power in hydrodynamic simulations is increased at smaller scales by radiative cooling and star formation, but is also broadly decreased by feedback processes inhibiting the clustering of matter \citep{Chisari_2018_baryonsPk, Chisari_2019, van_Daalen_2020, Delgado_2023_powerspec}. At larger scales in particular, feedback appears to play an important role in decreasing power, which has been supported by observations showing that stellar \citep{Lynds_1963_SNobs, Madau_1996_IGM, Martin_1998_GW, Pettini_2001_GW} and AGN \citep{Feruglio_2010_agn_feedback, Sturm_2011_agnobs, Greene_2012_AGN, Fabian_2012_agn_review, Cicone_2014_agnobs} feedback-driven outflows are capable of ejecting gas significant distances away from dark matter haloes. Though not an exhaustive list, the above effects alone can significantly alter the distribution of matter as compared to an $N$-body simulation, which further complicates efforts to account for the effects of baryons in such simulations.

One strategy to illuminate these complex feedback processes and to perhaps bypass the need to tightly constrain them is being carried out by the Cosmology and Astrophysics with MachinE Learning Simulations (CAMELS) project\footnote{\url{https://www.camel-simulations.org/}} \citep{Villaescusa_Navarro_2021}. CAMELS contains thousands of hydrodynamic and $N$-body simulations with wide ranging variations of cosmological and subgrid feedback parameters. Using the large library of simulations, CAMELS data have been used to account for these uncertain feedback processes in a variety of ways. Promising results have arisen from attempts to predict cosmological parameters while marginalizing over astrophysical effects \citep{Villaescusa-Navarro_2020_marginalization_example, Villaescusa-Navarro_2021_marginalization_fields2, Villaescusa-Navarro_2021_marginalization_fields, Villanueva-Domingo_2022_marginalizing_cosmicgraphs, Shao_2022_marginalization_subhaloes, Perez_2022_camelsSAM, deSanti_2023_fieldlevelgalaxies}, estimate the mass of dark matter haloes from baryonic properties \citep{Villanueva-Domingo_2021_marginalizing_milkyway, Villanueva-Domingo_2022_marginalizing_haloes} constrain subgrid feedback parameters, \citep{Thiele_2022_paramconstraint, Moser_2022_camelsCGM, Tillman_2023_lyalpha, Pandey_2023_powerspec}, search for other summary statistics that may contain valuable cosmological information \citep{Nicola_2022_sumstat, Villaescusa-Navarro_2022_OneGalaxy}, reduce the scatter of scaling relations \citep{Wadekar_2022_marginalizing, Wadekar_2022_scaling}, and more.

The hydrodynamic simulations in CAMELS include three different galaxy formation models: IllustrisTNG \citep{weinberger_2017_TNGpaper, Pillepich_2018_TNGpaper}, \textsc{SIMBA} \citep{SIMBA_Dave_2019}, and \textsc{ASTRID} \citep{bird_2022_ASTRIDgalaxys, ni_2022_ASTRIDSMBH}. In CAMELS, the strengths of SNe and AGN feedback parameters as prescribed by the respective models are varied, which allows for systematic analysis of the effects that these processes have on, for example, the cosmic star formation rate history, the galaxy stellar mass function, and the large-scale distribution of matter as a whole. 

In this work, we take advantage of the systematic model variations in CAMELS to quantify how far both dark and baryonic matter spread apart as a function of cosmological and feedback parameters by means of the Lagrangian matter {\it spread metric}. The spread metric was introduced in \cite{Borrow_2020} and was used to quantify the redistribution of matter in the \textsc{SIMBA} cosmological simulation. It was shown that 40\% of the baryonic content of the simulated volume can spread more than 1\,Mpc/$h$ away from the initial neighboring dark matter distribution owing to the impact of large-scale AGN jets in \textsc{SIMBA} (see also \citealt{SIMBA_Dave_2019}; \citealt{Christiansen_2020_SIMBA}). Here, we present a detailed analysis of cosmological baryon spread including thousands of galaxy formation model variations in CAMELS, including the \textsc{SIMBA}, IllustrisTNG and \textsc{ASTRID} implementations. Additionally, because feedback suppresses the matter power spectrum on large scales, we extend this analysis to investigate how the spreading of baryons correlates with the impact of feedback on the power spectrum.

This paper is organized as follows: In Section~\ref{Methodology}, we describe the CAMELS project, the datasets used, the spread metric, and other analysis techniques. In Section~\ref{Results}, we describe the results of analyzing the spread of dark matter and baryons in the \textsc{SIMBA} suite, as well as the correlation between cosmological baryonic spread and the total matter power spectrum. We also extend this analysis to simulations from the IllustrisTNG and \textsc{ASTRID} suites. In Section~\ref{Discussion}, we discuss the significance of these results in the context of the current state of the field. Finally, in Section~\ref{Conclusions}, we summarize the conclusions of this work.

\section{Methodology}
For this work, we focus first on presenting a detailed study of cosmological matter spread using the \textsc{SIMBA} simulation suite in CAMELS, which we will then compare to the IllustrisTNG and \textsc{ASTRID} simulation suites. Our simulations and relevant analysis techniques are described below.
\label{Methodology}
\subsection{CAMELS Simulations}
CAMELS \citep{Villaescusa_Navarro_2021} is a collection of 5516 (magneto)hydrodynamic cosmological simulations with subgrid physics from the IllustrisTNG, \textsc{SIMBA}, and \textsc{ASTRID} galaxy formation models, and 5164 $N$-body simulations with matching initial conditions, each with a comoving volume of $(\text{25\,Mpc}/h)^{3}$ containing $256^{3}$ dark matter particles evolving from $z = 127$ to present day.
Dark matter particles have a mass of $6.49 \times 10^{7} (\Omega_{\rm{m}} - \Omega_{\rm{b}})/0.251h^{-1}\rm{M_{\odot}}$. Hydrodynamic simulations contain an additional $256^{3}$ gas particles (which may form stars and black holes) with an initial mass of $1.27 \times 10^{7} \Omega_{\rm{b}}/0.049 h^{-1}\rm{M_{\odot}}$. The (sub)halo catalogs used in this work are generated with {\sc SUBFIND} \citep{Springel_2001_SUBFIND}.

The \textsc{SIMBA} galaxy formation model builds on the MUFASA model \citep{mufasa_Dave_2016}, and uses the ``Meshless Finite Mass'' mode of the $N$-body and hydrodynamics code GIZMO \citep{gizmo_Hopkins_2015}. The gravitational dynamics is solved with a Tree-PM method adapted from the GADGET-III code \citep{gadget_Springle_2005}. Radiative cooling and photoionization are implemented using Grackle-3.1 \citep{grackle_Smith_2016}. Stellar feedback is modeled similar to that of MUFASA: two-phase galactic winds with 30\% of wind particles being ejected with a temperature set by the difference in SNe energy and wind kinetic energy. As an update from MUFASA, the mass loading factor and velocity of galactic winds scale with stellar mass following the relations found from FIRE zoom-in simulations \citep{Muratov_FIRE_paper, DAA_2017}. Supermassive black hole (SMBH) growth is implemented in two phases, where the gravitational torque accretion model \citep{hopkins_and_quataert_2011, DAA_2013, DAA_2015, DAA_2017_torqueBHaccretion} is used for cold gas and the Bondi accretion model \citep{Bondi_1952} is used for hot gas. Feedback from AGN is comprised of mechanical quasar-mode winds and high-speed collimated jets at fixed momentum flux following \citet{DAA_2017_torqueBHaccretion} and X-ray feedback following \citet{Choi_2012}. For a more thorough description of \textsc{SIMBA}, see \cite{SIMBA_Dave_2019}.

The IllustrisTNG galaxy formation model builds on the Illustris model \citep{Genel_2014_illustris, Vogelsberger_2014_illustris2}, and uses the Arepo code \citep{springel_2010_arepo} with the Tree-PM method to solve the equations of gravity and a Voronoi moving-mesh method to solve for magnetohydrodynamics. Radiative cooling follows \cite{katz_1996_tngcooling}, \cite{wiersma_2009_tngcooling} \cite{Rahmati_2013_Hshielding}. Stellar feedback galactic winds follow a kinetic scheme based on \cite{springel_hernquist_2003_stellarfeedback} in which particles are stochastically and isotropically ejected from star-forming gas. The SMBH model builds upon \cite{gadget_Springle_2005}, \cite{sijacki_2007_AGNmodel}, and \cite{Vogelsberger_2013_cosmosim}, with SMBH mergers occurring when SMBH particles enter each other's ``feedback spheres,'' and with gas accretion following the Eddington rate-limited Bondi parameterization \citep{Bondi_1952}. AGN feedback is implemented in three modes: thermal, kinetic, and radiative. The high accretion (thermal) mode injects thermal energy at a rate proportional to the mass accretion rate into a ``feedback sphere'' around the SMBH, while the low accretion (kinetic) mode accumulates energy over time and injects kinetic energy in a random direction into the feedback sphere when a total amount of energy since last injection is accumulated. The radiative component is always active, and adds the SMBH's radiation flux to the cosmic ionizing background. IllustrisTNG is fully described in \citet{weinberger_2017_TNGpaper} and \citet{Pillepich_2018_TNGpaper}.

The \textsc{ASTRID} galaxy formation model is implemented in the MP-Gadget simulation code (expanded from GADGET-III), using the Tree-PM method to solve for gravity and the Smoothed Particle Hydrodynamics (SPH) method. Radiative cooling and heating are modeled from \cite{katz_1996_tngcooling}, \cite{Vogelsberger_2013_cosmosim}, \cite{Faucher_2020_ionizingbackground}, and \cite{Rahmati_2013_Hshielding}. Stellar feedback is implemented kinetically, with wind particles sourced from newly-formed star particles. The CAMELS version of \textsc{ASTRID} \citep{Ni_2023_CAMELSastrid} uses the SMBH model adapted from the ASTRID production run \cite{bird_2022_ASTRIDgalaxys,ni_2022_ASTRIDSMBH}, without explicit modelling of the BH dynamical friction, and is extended to include a two-mode SMBH feedback implementation using thermal or kinetic energy injection depending on the Eddington ratio of the SMBH accretion rate. The low accretion rate (kinetic) model follows \cite{weinberger_2017_TNGpaper} but with different parameter values. The high accretion rate (thermal) model injects 5\% of the SMBH's radiation energy into the surrounding gas within a spherical region with radius twice that of the SPH kernel. \textsc{ASTRID} is fully described in \cite{bird_2022_ASTRIDgalaxys} and \cite{ni_2022_ASTRIDSMBH}.

Throughout the CAMELS suites, cosmological and feedback parameters are varied in different ways across four sets of simulations. We first focus on variations of two cosmological parameters ($\Omega_{\rm{m}}$ and $\sigma_{8}$ with fixed $\Omega_{\rm{b}} = 0.049$), two parameters governing SNe feedback ($A_{\rm{SN1}}$ and $A_{\rm{SN2}}$), and two parameters governing AGN feedback ($A_{\rm{AGN1}}$ and $A_{\rm{AGN2}}$). We explore the implications of varying these parameters in the \textsc{SIMBA} model and how they affect the spreading of matter. In \textsc{SIMBA}, the feedback parameters represent the following quantities:
\begin{itemize}
  \item $A_{\rm{SN1}}$ - Mass loading factor of galactic winds.
  \item $A_{\rm{SN2}}$ - Speed of galactic winds.
  \item $A_{\rm{AGN1}}$ - Momentum flux of quasar and jet-mode feedback.
  \item $A_{\rm{AGN2}}$ - Speed of jet-mode feedback.
\end{itemize}
Feedback parameters are simply normalizations of feedback strength relative to the original feedback models (e.g. a \textsc{SIMBA} simulation with $A_{\rm{SN2}}$ = 2 will have twice the galactic wind speed as the original \textsc{SIMBA} subgrid model). Fiducial values of these six parameters are taken to be $\Omega_{\rm{m}}$ = 0.3, $\sigma_{8}$ = 0.8, and $A_{\rm{SN1}}$ = $A_{\rm{SN2}}$ = $A_{\rm{AGN1}}$ = $A_{\rm{AGN2}}$ = 1. Each parameter variation range is as follows: 0.1 $\le$ $\Omega_{\rm{m}}$ $\le$ 0.5; 0.6 $\le$ $\sigma_{8}$ $\le$ 1.0; 0.25 $\le$ ($A_{\rm{SN1}}, A_{\rm{AGN1}}$) $\le$ 4.00; and 0.5 $\le$ ($A_{\rm{SN2}}, A_{\rm{AGN2}}$) $\le$ 2.0. After a focused analysis of the effects variations on these six parameters have on the spreading of dark matter and gas in SIMBA, we will explore a broader range of up to 28 parameter variations (see \citealt{Ni_2023_CAMELSastrid} for descriptions of all varied parameters) comparing the baryon spreading in SIMBA, IllustrisTNG, and ASTRID.

For each simulation, the total matter power spectrum $P(k)$ is calculated following \cite{Villaescusa_Navarro_2021}. The masses for every particle type (dark matter, gas, stars, and black holes) are placed in a $512^{3}$ voxel grid, which is Fourier transformed by averaging over $k$ bins. The bin width is the fundamental frequency, $k_{F} = 2\pi/L$, where $L$ is the length of the simulated box, $\text{25\,Mpc}/h$.

\subsection{Datasets}
This analysis uses several different datasets within CAMELS. Sections \ref{DM Results} through \ref{Pk Results} all use the six parameter 1P set, CV set, and LH set from the \textsc{SIMBA} suite presented in \cite{Villaescusa_Navarro_2021}, while section \ref{Comparisons} uses the full 28 parameter 1P set from the \textsc{SIMBA} suite, the 1P and LH sets from the \textsc{ASTRID} suite, and the 28 parameter 1P and SB28 sets from the IllustrisTNG suite presented in \cite{Ni_2023_CAMELSastrid}. A short description for each simulation set is given below:

\begin{itemize}
    \item The 1P (``one parameter'') set contains simulations with the same initial conditions (random seed) but varying one parameter at a time. One simulation uses all fiducial parameter values while the remaining simulations correspond to variations of up to 28 parameters above and below their fiducial value while all others are held constant. The six parameters described above each have 10 variations (hereafter referred to as the six parameter 1P set), while the remaining 22 parameters each have 4 variations. The full list and descriptions of all 28 parameters can be found in \cite{Ni_2023_CAMELSastrid}. The parameter variation spacing is linear for cosmological parameters and logarithmic for feedback parameters. The corresponding $N$-body 1P set contains 21 simulations varying only $\Omega_{\rm{m}}$ and $\sigma_{8}$. The 1P set is designed for determining the effects of specific parameters on various quantities such as the matter spread.
    \item The CV (``cosmic variance'') set contains 27 simulations with the same fiducial parameters, but different initial conditions. The CV set is used to quantify the effects of cosmic variance on observables, which is important given that the simulated volumes in CAMELS are small and not representative of the Universe as a whole. 
    \item The LH (``Latin Hypercube'') set contains 1000 simulations each with different initial conditions, and with near-random parameters selected from a Latin hypercube. The main goal of the LH set is to train machine learning algorithms to make predictions given cosmological and astrophysical inputs, account for cosmic variance, and marginalize over baryonic effects.
    \item The SB28 ("Sobol Sequence") set is unique to IllustrisTNG and contains 1024 simulations with 28 parameters quasi-randomly selected from a Sobol sequence designed for machine learning applications in an expanded parameter space \citep{Ni_2023_CAMELSastrid}.
\end{itemize}

\subsection{The Spread Metric}
One way to quantify the decoupling of the baryonic and dark matter components is with the cosmological {\it spread metric} \citep{Borrow_2020}. The spread metric for a particle (either gas or dark matter) is defined as the final distance at $z=0$ between that particle and the dark matter particle it was nearest to in the initial conditions. For the spread of gas in IllustrisTNG, we use tracer particles that are designed to accurately follow the flow of gas (see \citealt{genel_2013_tracers} for a full description) rather than tracking gas cells. For \textsc{SIMBA} and \textsc{ASTRID}, gas particles that turn into stars are not included in this analysis. The method for calculating the spread of a particle (which is the same for both gas and dark matter) is as follows:

\begin{enumerate}
  \item In the initial conditions, find the particle's nearest dark matter neighbor by computing the distance to all dark matter particles.
  \item Store the IDs of these two neighboring particles.
  \item By ID matching, find these particles in the $z=0$ snapshot and compute the distance between them.
\end{enumerate}

Only the initial conditions and final state of the simulation are required; here we focus on the spread of matter from $z=127$ down to $z=0$. The distributions of dark matter and gas particles are identical at $z=127$ except for the systematic displacement of the staggered grids that define the initial conditions. The neighboring dark matter particle is used as a reference point in this metric due to the ambiguity involved in measuring absolute distances in an evolving universe, where the gas and dark matter components in large structures can drift owing to peculiar motions. The spread metric is thus a measure of displacement relative to the initial surrounding dark matter distribution normalizing out peculiar motions. In this work, we calculate the ``spread'' value for every gas and dark matter particle in all CAMELS simulations. The spread metric can also be computed for star particles, but here we focus on the spread of gas and refer to it as baryon spread interchangeably. With the spread output, particles may be selected and analyzed in various ways, such as investigating the spread of particles inside or outside of haloes, particles within a specified halo mass range, or particles within a specified spread percentile range. The median spread of particles for a given simulation can be computed to characterize with a single statistic the overall impact of baryonic effects on the distribution of matter.

\begin{figure}
	\includegraphics[width=\columnwidth]{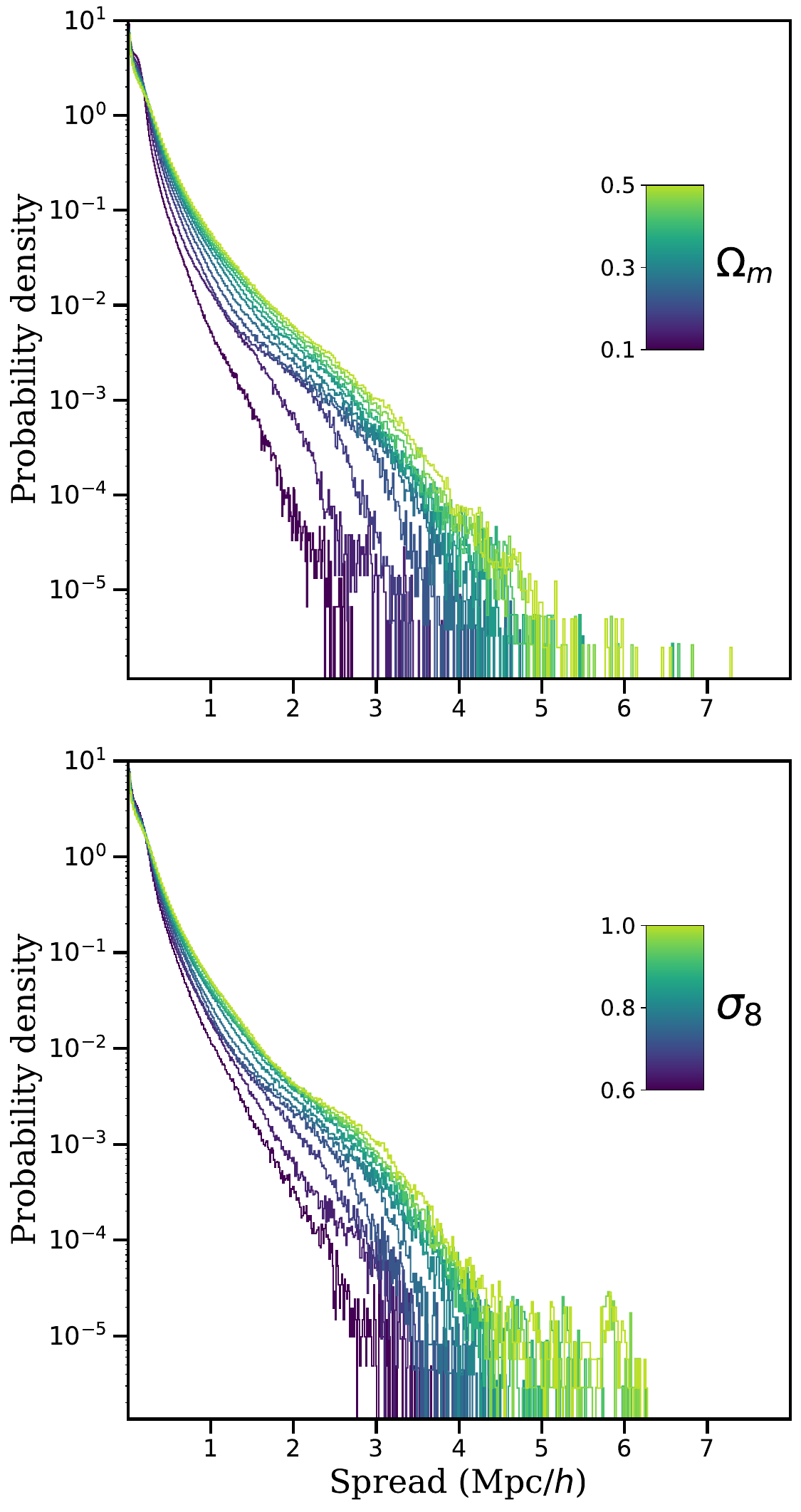}
    \caption{Distribution of spread distances for all dark matter particles in the 1P set $N$-body simulations at $z=0$. The top panel shows the spread distribution in simulations varying $\Omega_{\rm{m}}$ (with all other parameters held constant) and the bottom panel shows the impact of varying $\sigma_{8}$ alone. Color bars indicate the parameter value for each simulation. Dark matter spreads further from the initial neighboring distribution with increasing $\Omega_{\rm{m}}$ and $\sigma_{8}$.}
    \label{fig:dm_spread}
\end{figure}

\subsection{Symbolic Regression}
Symbolic regression is a machine learning technique capable of finding an analytic mathematical formula to relate an input and an output. A distinct advantage of symbolic regression over a more standard ``least squares'' regression is that the functional form does not need to be known ahead of time, and instead is constructed from a set of allowed operators. We use symbolic regression to find a function that can predict the impact of baryonic feedback on the matter power spectrum as a function of baryon spread and wave number $k$. We use the PySR package \citep{pysr} to train on SIMBA six parameter 1P set simulations that vary the four feedback parameters where the input is a 2D array with one axis consisting of median gas spread values for each simulation and the other consisting of 40 $k$ values between 0.5 and $10\,h\,\rm{Mpc^{-1}}$. The output is the (negative) change in total matter power spectrum between the hydrodynamic and $N$-body simulations, $-\Delta P/P = -(P_{\rm{Hydro}} - P_{\rm{Nbody}})/P_{\rm{Nbody}}$. The allowed operators are addition, multiplication, subtraction, division, exponential, inverse, absolute value, and square root. We use the loss function default to PySR, mean squared error:
\begin{equation}
    f_{\rm{loss}} = \frac{1}{N} \sum_{i}^{N}(\Delta P/P^{\rm{true}}_{i} - \Delta P/P^{\rm{predicted}}_{i})^{2},
\end{equation}
where N is the total number of data points. We use the measure of complexity default to PySR, which treats each operator with equal complexity. Finally, we use 100 iterations in each search.

\section{Results}
\label{Results}

In this section, we use the spread metric to analyze the redistribution of dark matter in $N$-body simulations and gas in SIMBA hydrodynamic simulations to evaluate how these change with varying cosmological parameters and (in the case of gas) feedback strength. We then investigate the correlation between the spread of gas and the impact of baryons on the total matter power spectrum. We conclude this section with comparisons between the baryon spread and impact on the matter power spectrum in \textsc{SIMBA}, IllustrisTNG, and \textsc{ASTRID}.

\subsection{Dynamical spread in $N$-body simulations}
\label{DM Results}
All particles are subject to the force of gravity and may spread owing to chaotic gravitational dynamics. We begin by analyzing the spreading of dark matter in $N$-body simulations, which will serve as control for our subsequent analysis of baryonic spread to understand the relative roles of gravity and baryonic physics on the redistribution of matter. 

\begin{figure}
	\includegraphics[width=\columnwidth]{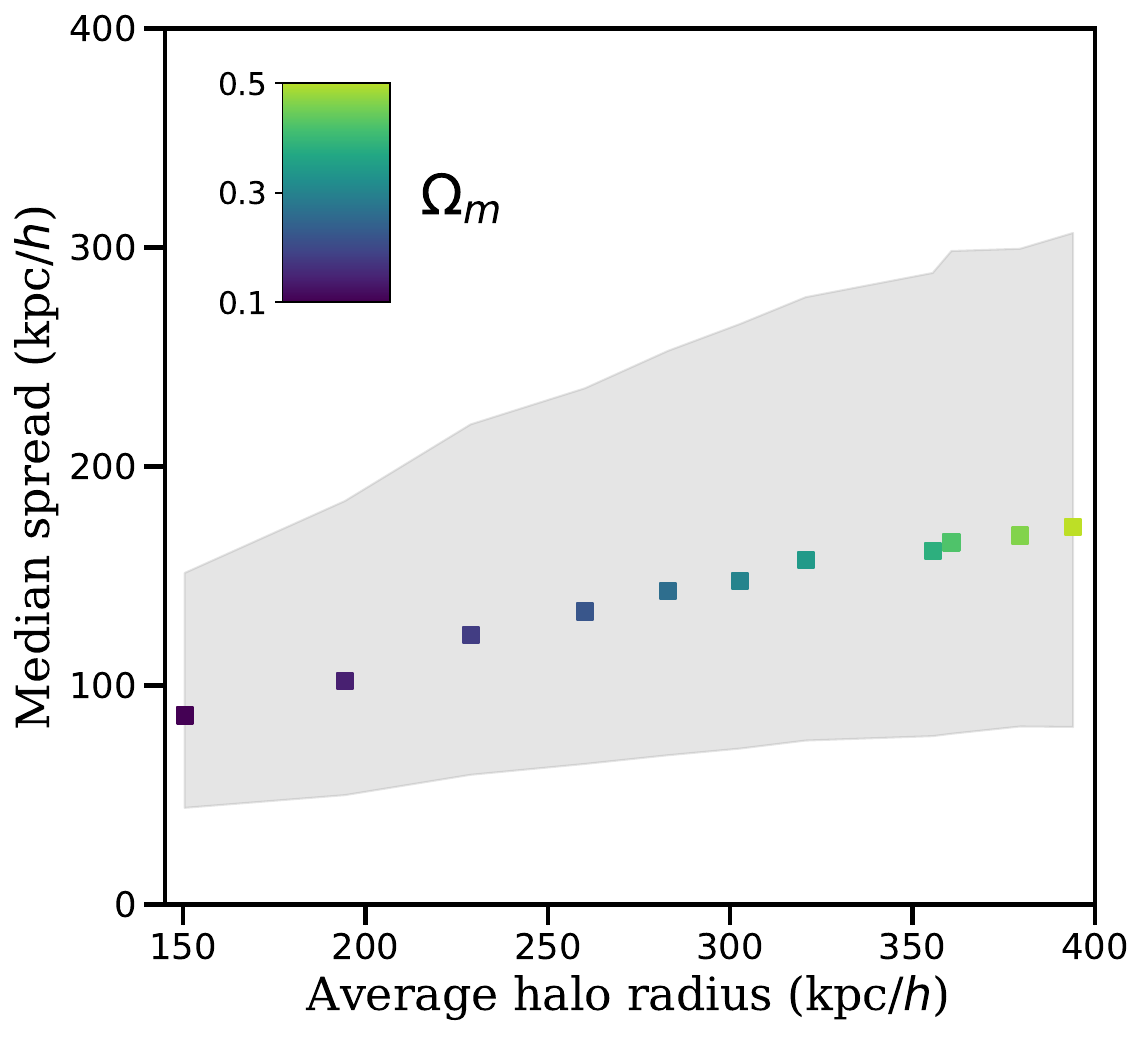}
    \caption{Median spread of dark matter particles in the 10 most massive haloes (excluding the most massive halo) as a function of the average virial radius among these haloes at $z=0$ in each of the $N$-body simulations varying $\Omega_{\rm{m}}$. The color bar indicates the value of $\Omega_{\rm{m}}$ for each simulation. As $\Omega_{\rm{m}}$ increases, haloes get larger and spreading increases.}
    \label{fig:dm_max_halo}
\end{figure}

\begin{figure}
	\includegraphics[width=\columnwidth]{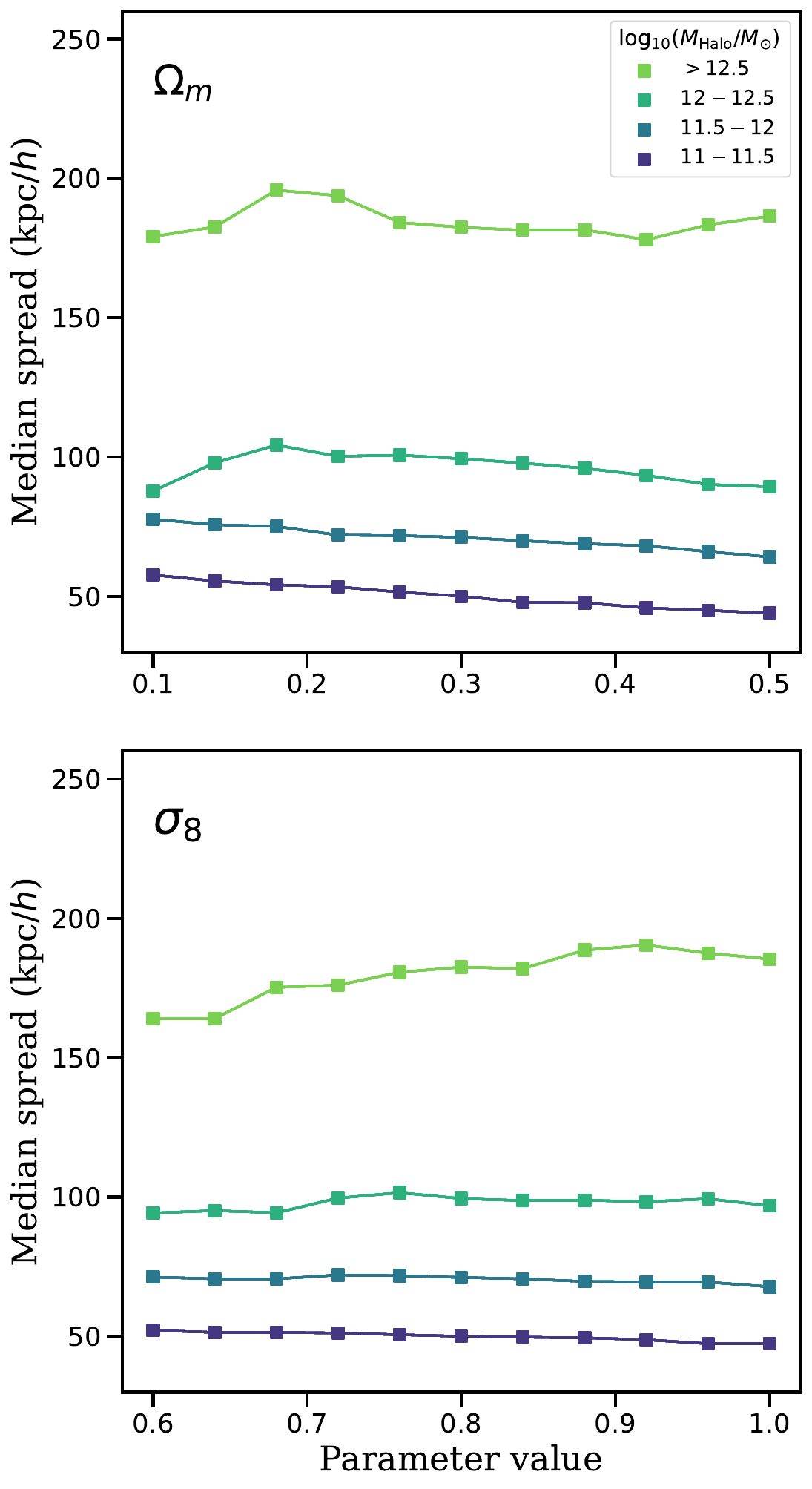}
    \caption{Median spread of dark matter particles inside of halos of selected mass ranges as a function of $\Omega_{\rm{m}}$ (top) and $\sigma_{8}$ (bottom) in each of the $N$-body simulations at $z=0$. Color represents the halo mass range as described by the legend in the top panel. Binning by halo mass removes most of the dependence of spread on cosmological parameters and shows a clear positive correlation between halo mass and median spread.}
    \label{fig:dm_halo_masses}
\end{figure}

Using the $N$-body 1P set, we can investigate how the spreading of dark matter depends on cosmological parameters. Figure \ref{fig:dm_spread} shows the distribution of the spread of all dark matter particles in each simulation as the cosmological parameters $\Omega_{\rm{m}}$ and $\sigma_{8}$ are varied. Each line corresponds to a different simulation and the color denotes the parameter value. We see here that higher parameter values for $\Omega_{\rm{m}}$ and $\sigma_{8}$ correspond to greater spread of dark matter (more particles are spreading farther from their initial neighbors). At the greatest values of $\Omega_{\rm{m}}$, some particles are spreading more than 7\,Mpc/$h$, whereas the farthest spread particles in the lowest $\Omega_{\rm{m}}$ run spread less than 3\,Mpc/$h$. Variations in $\sigma_{8}$ show a slightly tighter distribution, with the farthest spread particles in the highest $\sigma_{8}$ run spreading just over 6\,Mpc/$h$, and around 3\,Mpc/$h$ in the lowest $\sigma_{8}$ run. 

\begin{figure*}
	\includegraphics[width=\textwidth]{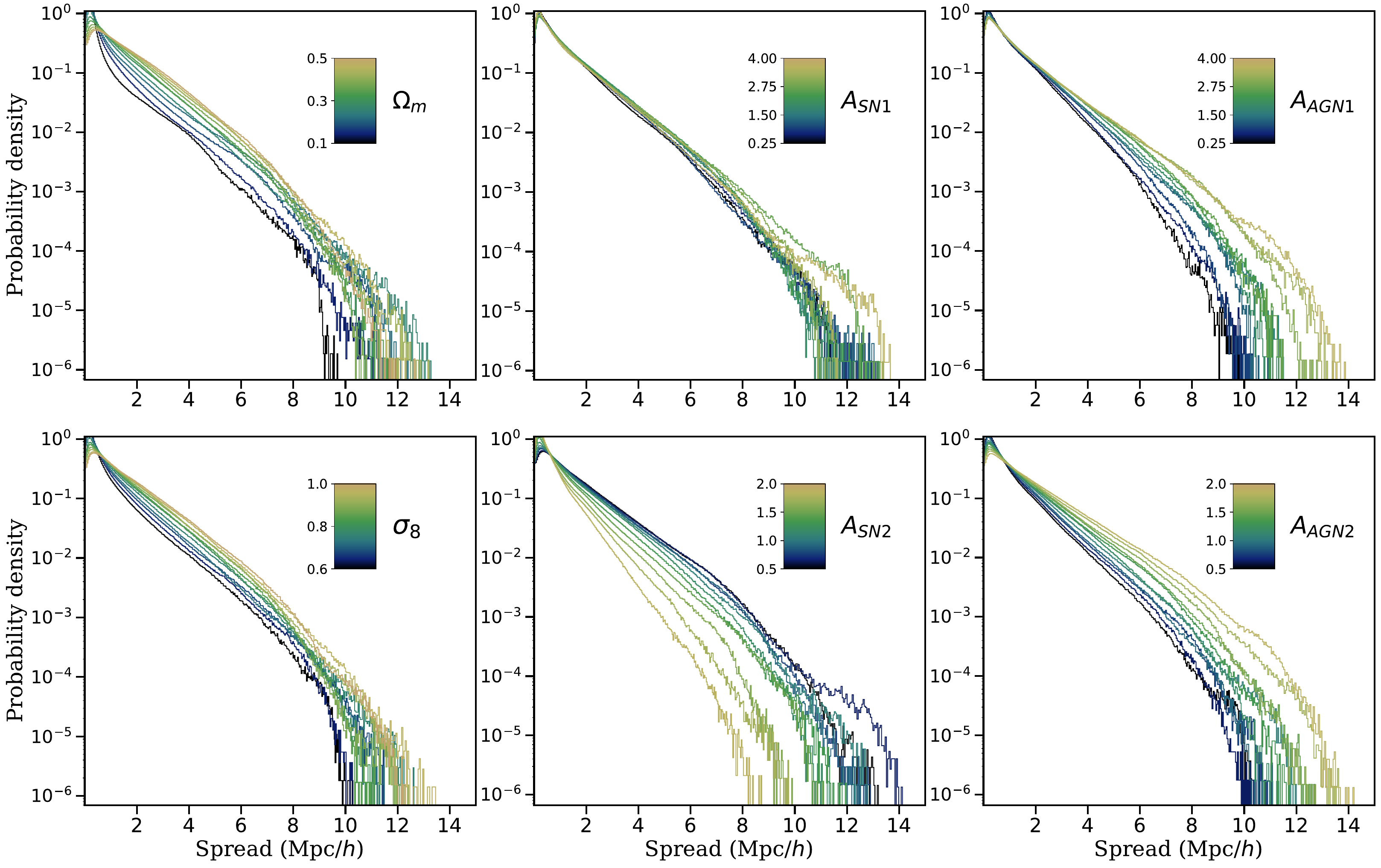}
    \caption{Distribution of spread distances for gas particles in the six parameter \textsc{SIMBA} 1P set at $z=0$. Each panel contains distributions corresponding to one simulation run in which all other parameters are held constant at fiducial values. Color bars indicate parameter value for each simulation. For both cosmological parameters and AGN parameters, spread is clearly positively correlated, while it shows a minor positive correlation with $A_{\rm{SN1}}$ (mass loading) and a strong negative correlation with $A_{\rm{SN2}}$ (wind speed).}
    \label{fig:gas_spread}
\end{figure*}

Dark matter halos represent gravitational potential wells where chaotic orbits can make the trajectories of dark matter particle neighbors diverge over time. The dark matter particles that spread the most are then likely to reside around the most massive structures in the simulation. It is thus informative to explore the relationship between spread, parameter value, and halo size. Figure \ref{fig:dm_max_halo} shows the median spread of dark matter particles from the 10 most massive haloes (but excluding the most massive as it is often significantly larger than the others) as a function of their average virial radius for simulations varying $\Omega_{\rm{m}}$, with the shaded region representing the 25th to 75th percentile of spread. This depicts a clear positive (sub-linear) correlation between $\Omega_{\rm{m}}$ and the average virial radius, which increases from 150 kpc/$h$ to 395 kpc/$h$ with $\Omega_{\rm{m}} = 0.1 \rightarrow 0.5$ and also correlates with an overall increase in the spread of particles. This suggests that increasing $\Omega_{\rm{m}}$ and $\sigma_{8}$ yield wider spreading of matter by increasing the mass and abundance of massive halos (see \citealt{Villaescusa_Navarro_2021}).

To investigate further, Figure \ref{fig:dm_halo_masses} shows the median spread of particles residing within haloes of specified mass ranges as a function of parameter value. Regardless of cosmological parameter variation, particles in larger haloes spread farther. When looking at haloes of the same mass, the values of $\Omega_{\rm{m}}$ and $\sigma_{8}$ have little impact on the median spread of dark matter. Dark matter spread does, however, slightly decrease at fixed halo mass with increasing $\Omega_{\rm{m}}$, which may be explained by an increase in halo concentration (see Section~\ref{Discussion}). The clear relationship between halo mass and dark matter spread provides support for the increased dark matter spread at higher $\Omega_{\rm{m}}$ and $\sigma_{8}$ simply reflecting the formation of more massive haloes. This relationship is further supported by the full spread distribution of particles in these haloes, which shows a clear dependence on halo mass at fixed cosmology (see Appendix \ref{appendix}).

\subsection{Baryonic spread in cosmological hydrodynamic simulations}
\label{Baryon Results}

The spreading of gas becomes more complicated with the addition of radiative cooling, hydrodynamic forces, and galaxy formation feedback in cosmological hydrodynamic simulations. Figure \ref{fig:gas_spread} shows the distribution of spread distances for all gas particles in each hydrodynamic simulation in the six parameter \textsc{SIMBA} 1P set, which now includes more simulations varying feedback parameters. As expected, the distance to which baryons spread relative to the initial neighboring dark matter distribution increases with higher values of $\Omega_{\rm{m}}$ and $\sigma_{8}$, reflecting the formation of more massive halos and the spread of matter owing to chaotic gravitational dynamics as seen for $N$-body simulations (Figure \ref{fig:dm_spread}). However, there is a notable difference in the spread of baryons compared to dark matter, with gas particles spreading up to $\sim$11\,Mpc/$h$ (compared to $\sim$4.5\,Mpc/$h$ for dark matter) in the fiducial simulation. 

Increasing the strength of AGN feedback systematically increases the spread of gas, with the maximum distance reached varying from $\sim$10--14\,Mpc/$h$ when increasing either $A_{\rm{AGN1}}$ or $A_{\rm{AGN2}}$ from their minimum to maximum parameter values explored here. 
In contrast, increasing the strength of SNe feedback shows complex results. Greater mass loading of SNe-driven winds ($A_{\rm{SN1}}$) yields an unclear, but perhaps minor positive correlation with spread, while greater speed of SNe-driven winds ($A_{\rm{SN2}}$) shows a strong negative correlation. This indicates that in \textsc{SIMBA}, overly efficient SNe feedback reduces the spreading of gas to large scales, which can be explained by driving a net reduction of energy injection by AGN feedback due to the evacuation of gas from central areas (see Section \ref{Discussion}). We note here that while these distances are comparable to the size of the simulated box in CAMELS, \cite{Borrow_2020} found that the maximum gas spread distance in the fiducial 100\,Mpc/$h$ SIMBA simulation was within half the CAMELS boxsize and thus should not affect results.

\begin{figure}
	\includegraphics[width=\columnwidth]{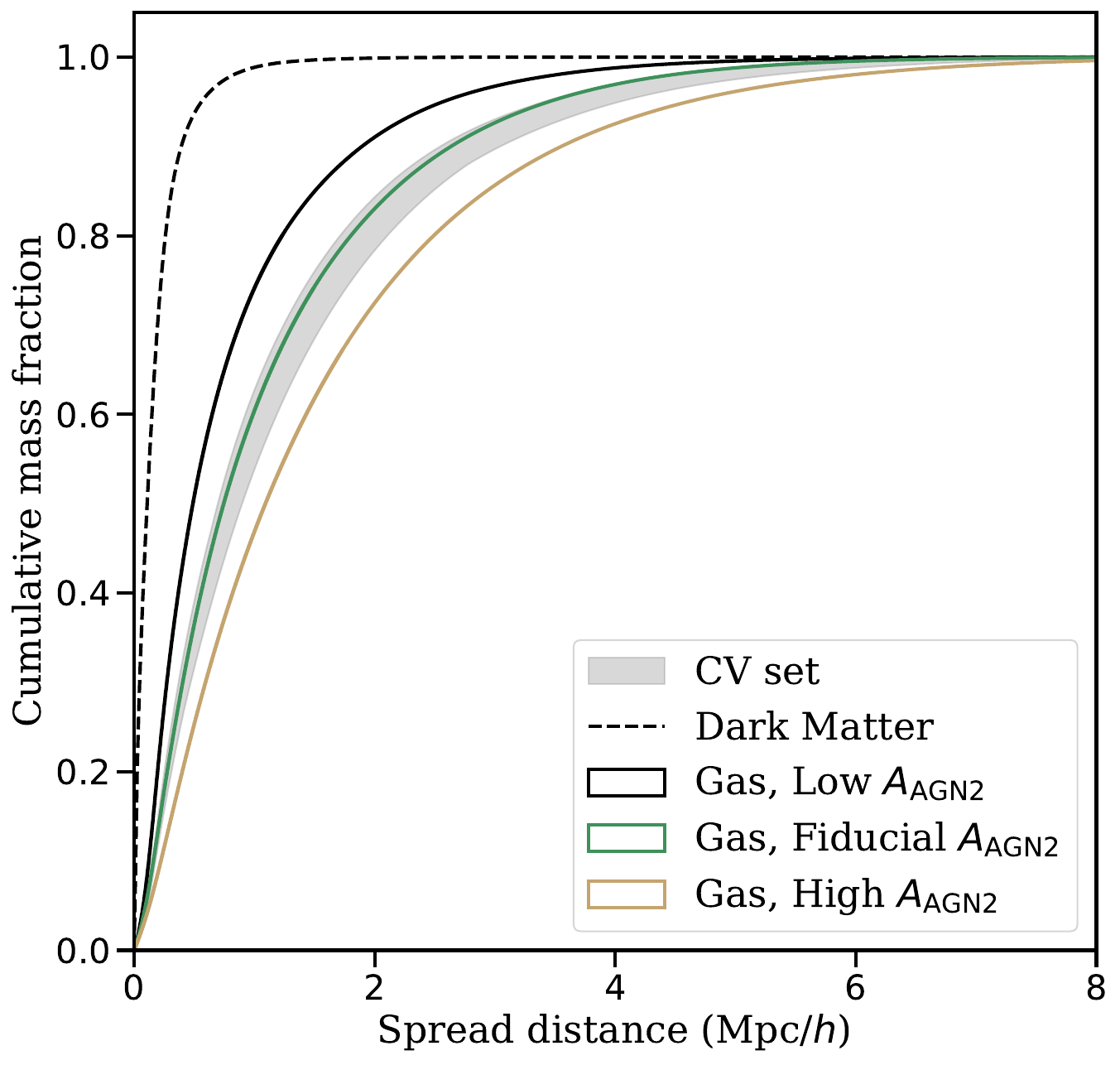}
    \caption{Cumulative mass distribution as a function of spread distance at $z=0$ for dark matter (dashed) and gas (green) particles in the fiducial \textsc{SIMBA} 1P simulation, for gas particles in the highest (brown) and lowest (black) $A_{\rm{AGN2}}$ (jet speed) simulations, and for gas particles in the full \textsc{SIMBA} CV set (the grey shaded area spans the full CV set distribution). In all cases, gas spreads much further than dark matter and increases with AGN jet speed.}
    \label{fig:cumulative_spread}
\end{figure}

Figure \ref{fig:cumulative_spread} compares the spread of dark matter and gas in different simulations in the form of cumulative distributions which more clearly show the amount of mass spreading up to a given distance. We compare the fiducial 1P simulation to the highest and lowest AGN jet speed models as well as the full spread distribution of the CV set. As expected, gas spreads significantly farther than dark matter (20\% of gas spreads farther than 1.8\,Mpc/$h$ compared to 0.26\,Mpc/$h$ for dark matter), and the spreading of gas increases with faster jet speeds (20\% of gas spreads farther than 1.23\,Mpc/$h$ with the lowest jet speed as compared to 2.5\,Mpc/$h$ with the highest jet speed).
The CV set shows the extent to which varying initial conditions can affect the large-scale spreading of material, which can be attributed to differences in the halo mass function and, in particular, the abundance of the most massive halos hosting SMBHs with powerful AGN jet feedback. In the CV set simulation with the least amount of gas spread, 20\% of gas spreads farther than 1.7\,Mpc/$h$, while in the simulation with the highest gas spread, 20\% of gas spreads farther than 2.1\,Mpc/$h$.

\begin{figure*}
	\includegraphics[width=\textwidth]{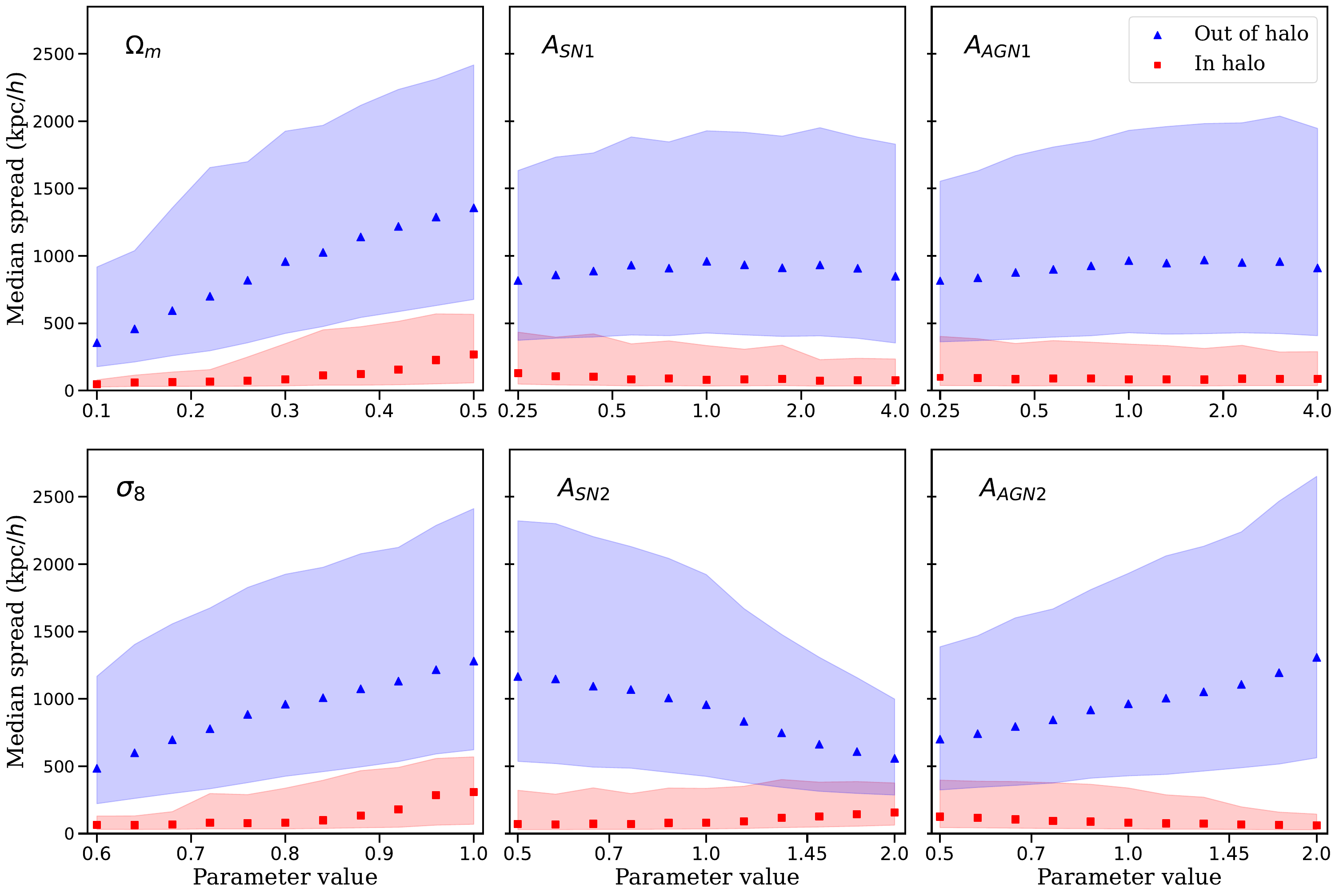}
    \caption{Cosmological spread of gas selected from halo Lagrangian regions at the initial conditions (i.e. gas particles whose nearest initial dark matter particle neighbor ends up inside of a halo at $z=0$) as a function of cosmological and feedback parameters in \textsc{SIMBA}. Median spreads are shown separately for gas that ends up inside of the corresponding halo (squares, red shading) or outside of any halo (triangles, blue shading) at the end of the simulation ($z=0$). Each panel corresponds to one individual parameter variation, as indicated. Shaded regions denote the 25th (lower) and 75th (upper) percentile of spread. The median spread of gas ejected from halo Lagrangian regions is significantly larger than for gas that remains inside of haloes.}
    \label{fig:gas_spread_haloes}
\end{figure*}

For gas, we expect the majority of spreading to occur around haloes, where galaxies act as sources of stellar and AGN feedback powering the ejection of gas to large distances. However, the spreading of gas in haloes becomes complex due to competing effects: feedback pushing gas outward and the ability for gas to radiate away energy (radiative cooling) and fall to lower radii in the gravitational potential well of dark matter haloes. Figure \ref{fig:gas_spread_haloes} depicts this dichotomy by quantifying the spread of gas initially located inside of the Lagrangian regions of $z=0$ halos at the initial conditions. For each halo at $z=0$, we define its Lagrangian region by tracking the corresponding dark matter particles back to their location at the initial conditions. Gas particles are then defined to be in a Lagrangian region at the initial conditions if their nearest dark matter particle neighbor ends up in a halo at $z=0$ \citep{Borrow_2020}. All of the selected gas particles shown in Figure \ref{fig:gas_spread_haloes} were initially located inside of a Lagrangian region, meaning that their nearest dark matter neighbor particle is inside of a halo at $z=0$. We compute separately the median spread distance for Lagrangian region gas that ends up inside of halos (red squares) or outside of halos (blue triangles) at $z=0$, and we indicate the 25th to 75th percentile range as shaded regions. Gas particles from Lagrangian regions that end up outside of halos spread much farther than the gas particles that remain in a halo regardless of parameter variations, as expected. Additionally, the spread of gas outside of haloes has a stronger dependence on parameter variations than gas that remains inside of halos. The dichotomy of baryonic spread seen here shows that Lagrangian regions contain baryons that will spread very little (dense gas converting into stars in the cores of dark matter halos) and very far (gas ejected in high-speed AGN jets).

\begin{figure*}
	\includegraphics[width=\textwidth]{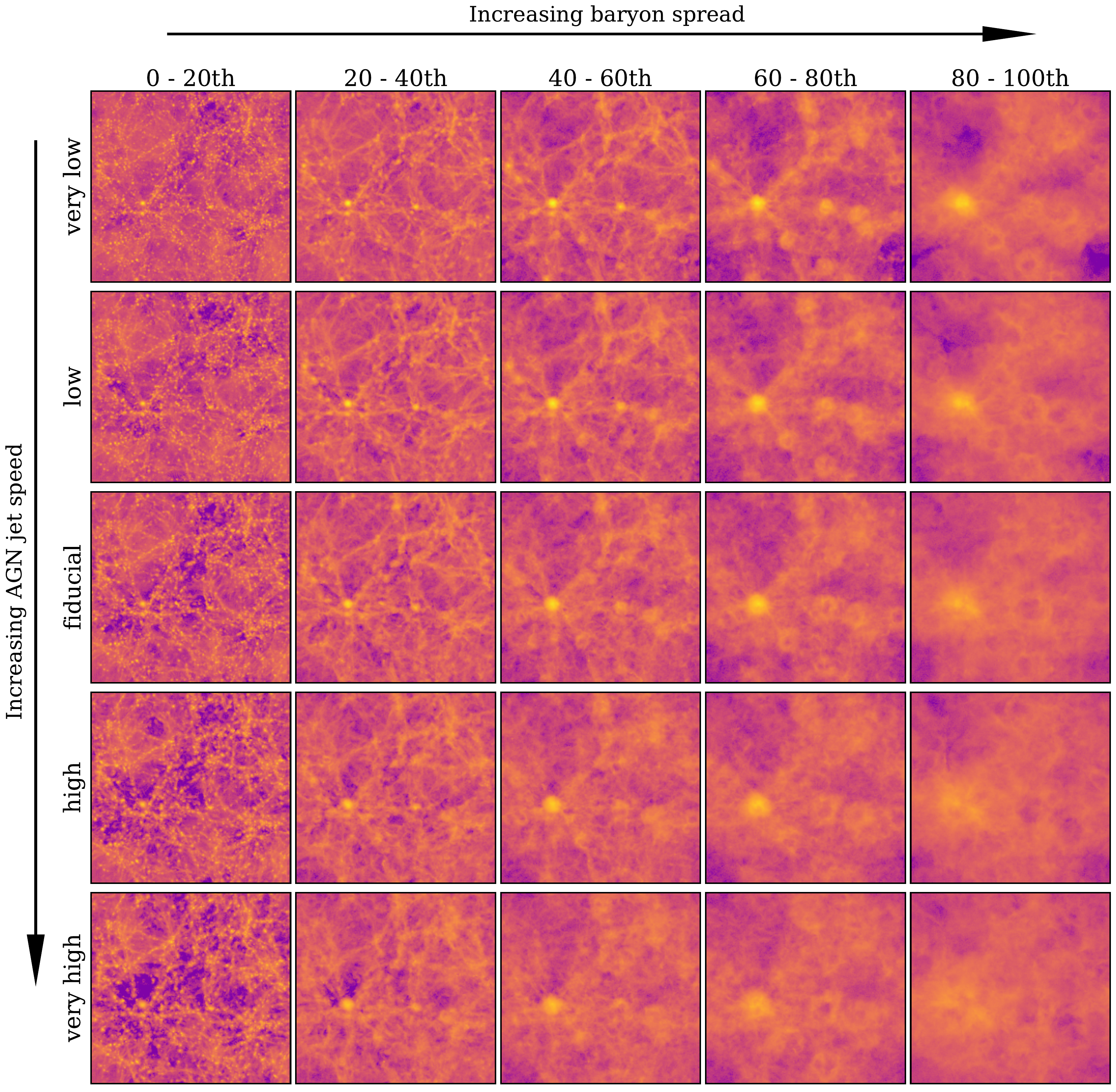}
    \caption{Spatial distribution of gas in a $(\text{25\,Mpc}/h)^{3}$ volume as a function of AGN jet speed ($A_{\rm{AGN2}}$; rows) and relative amount of spread at $z=0$ (columns) in \textsc{SIMBA}. Each panel in a given row is a 2D gas mass projection at $z=0$ containing 20\% of the baryonic content of the simulation, ranked by amount of spread from left (lowest spread percentile range) to right (highest spread percentile range). From top to bottom, simulations with progressively higher $A_{\rm{AGN2}}$ parameter values are shown. Color scale represents the mass density in the 2D projection and is logarithmic and identical in all panels. Particles of low spread percentile tend to be inside halos and filaments, while greatly-spread particles appear to reside in bubbles around halos and filaments.}
    \label{fig:gas_AGN2_25}
\end{figure*}

Figure \ref{fig:gas_AGN2_25} illustrates the spatial distribution of gas at $z=0$ that has spread by different amounts within a given simulated volume, comparing simulations with different AGN jet speed in \textsc{SIMBA}. Each panel represents a 2D mass projection of 20\% of the gas particles in the simulation, and thus each panel contains the same amount of mass. Each row corresponds to one simulation, and each column denotes the percentile of spread of the particles being plotted. The $A_{\rm{AGN2}}$ parameter (jet speed) is increased from top to bottom with values  $A_{\rm{AGN2}} \approx [0.5, 0.66, 1.0, 1.52, 2.0$]. In the low-spread panels (1st column), gas particles trace both the densest regions at the centers of haloes as well as a diffuse component far from the feedback generated by the most massive halos. As the spread increases, gas particles trace regions around massive haloes and filaments at increasingly large distances from halo centers. The contrast in the relative distributions of gas for different percentile ranges of spread is enhanced when increasing the AGN jet speed.

\subsection{Impact of baryon spread on the matter power spectrum}
\label{Pk Results}
\begin{figure}
	\includegraphics[width=\columnwidth]{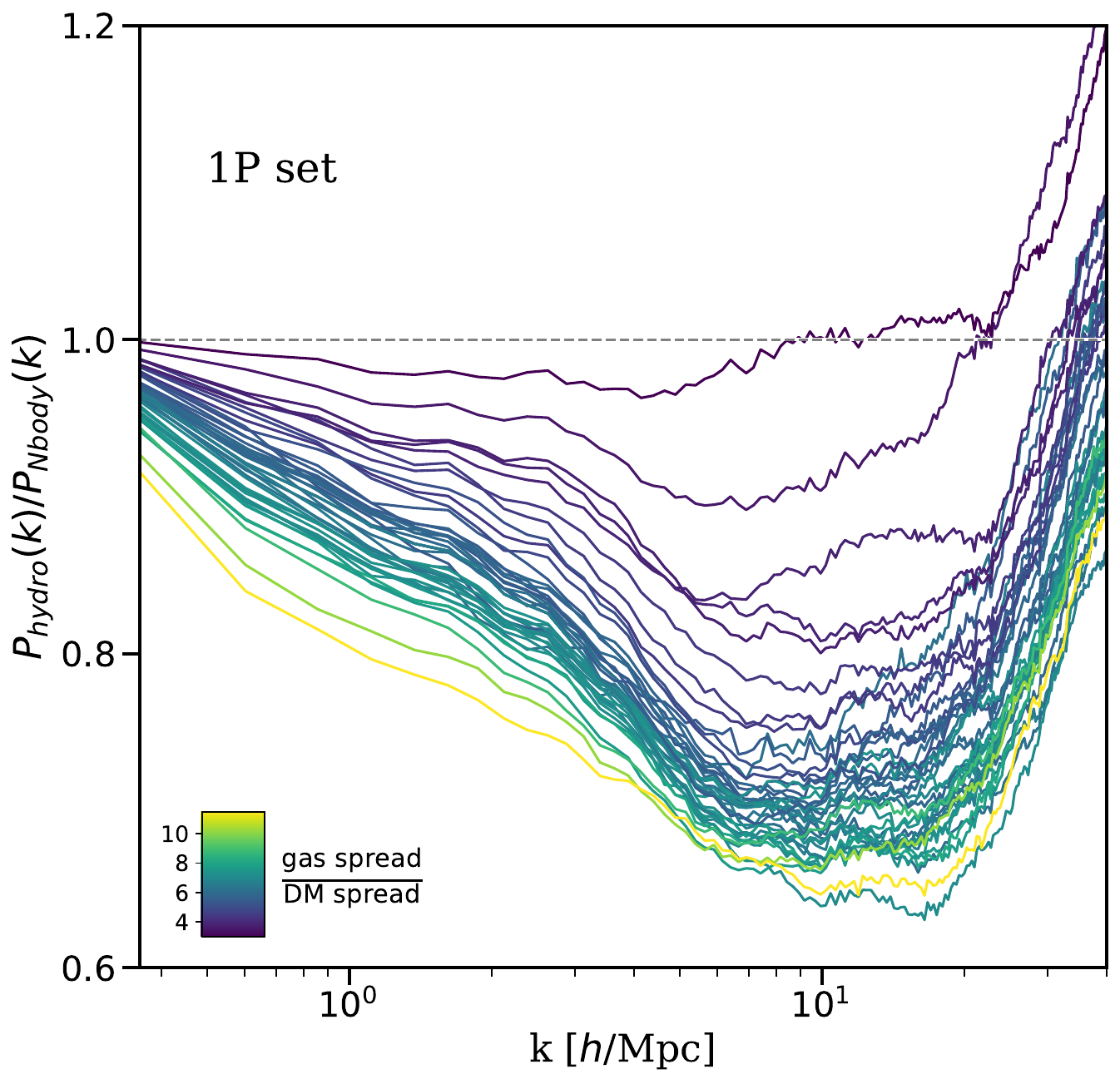}
    \caption{Ratio of hydrodynamic and $N$-body simulation power spectra for the 44 simulations from the six parameter \textsc{SIMBA} 1P set that vary feedback parameters (no cosmology variations), at $z=0$.
    Color scale corresponds to the median of gas particle spread divided by the initial neighboring dark matter particle spread for gas selected to have a dark matter neighbor in one of the five most massive haloes in the final snapshot. There exists a clear correlation between suppression of power on all scales and median normalized gas spread.}
    \label{fig:1P_PS}
\end{figure}

Using the hydrodynamic and $N$-body simulation pairs in CAMELS, we can investigate the impact of baryonic effects on the total matter power spectrum as a function of feedback strength, cosmic variance, and cosmology. Figure \ref{fig:1P_PS} shows the ratio of hydrodynamic and $N$-body power spectra for the six parameter \textsc{SIMBA} 1P set simulations that vary feedback parameters. If the distribution of matter was exactly the same in both simulations, the ratio of these power spectra would be $P_{\rm{hydro}}/P_{\rm{Nbody}} = 1$ at all values of $k$. Therefore, deviance from a ratio equaling 1.0 indicates baryonic impact on the matter power spectrum. At the smallest spatial scales, the ratio increases far above 1.0 due to radiative cooling of baryons and the formation of stars. On larger spatial scales, the ratio decreases below 1.0 due to feedback processes redistributing baryons and (to a much lesser extent) dark matter owing to back-reaction \citep{van_Daalen_2011, Chisari_2019}. To investigate the origin of these effects, we color-code the $P_{\rm{hydro}}/P_{\rm{Nbody}}$ lines based on a normalized measure of the amount of baryonic spread for each hydrodynamic simulation. In particular, we choose only gas particles that are in one of the five most massive Lagrangian regions in each simulation, as these regions generate the most feedback (our results are insensitive to the exact number of haloes considered). Additionally, we normalize the spread of each gas particle by dividing by the spread of its dark matter neighbor, partially mitigating the effect of varying cosmological parameters on the halo mass function and thus the (gravitational) dynamic spreading of matter. Lastly, we compute the median of this normalized baryonic spread for all selected gas particles to compare CAMELS simulations with different model parameters. For the six parameter SIMBA 1P set simulations varying feedback parameters (Figure \ref{fig:1P_PS}), we see a clear relationship between the impact of feedback on the clustering of matter and the spreading of gas particles across a range of scales, with greater spread correlating with greater reduction of power. This relationship continues to hold even at small scales ($k \sim 30\,h$\,Mpc$^{-1}$).

\begin{figure}
	\includegraphics[width=\columnwidth]{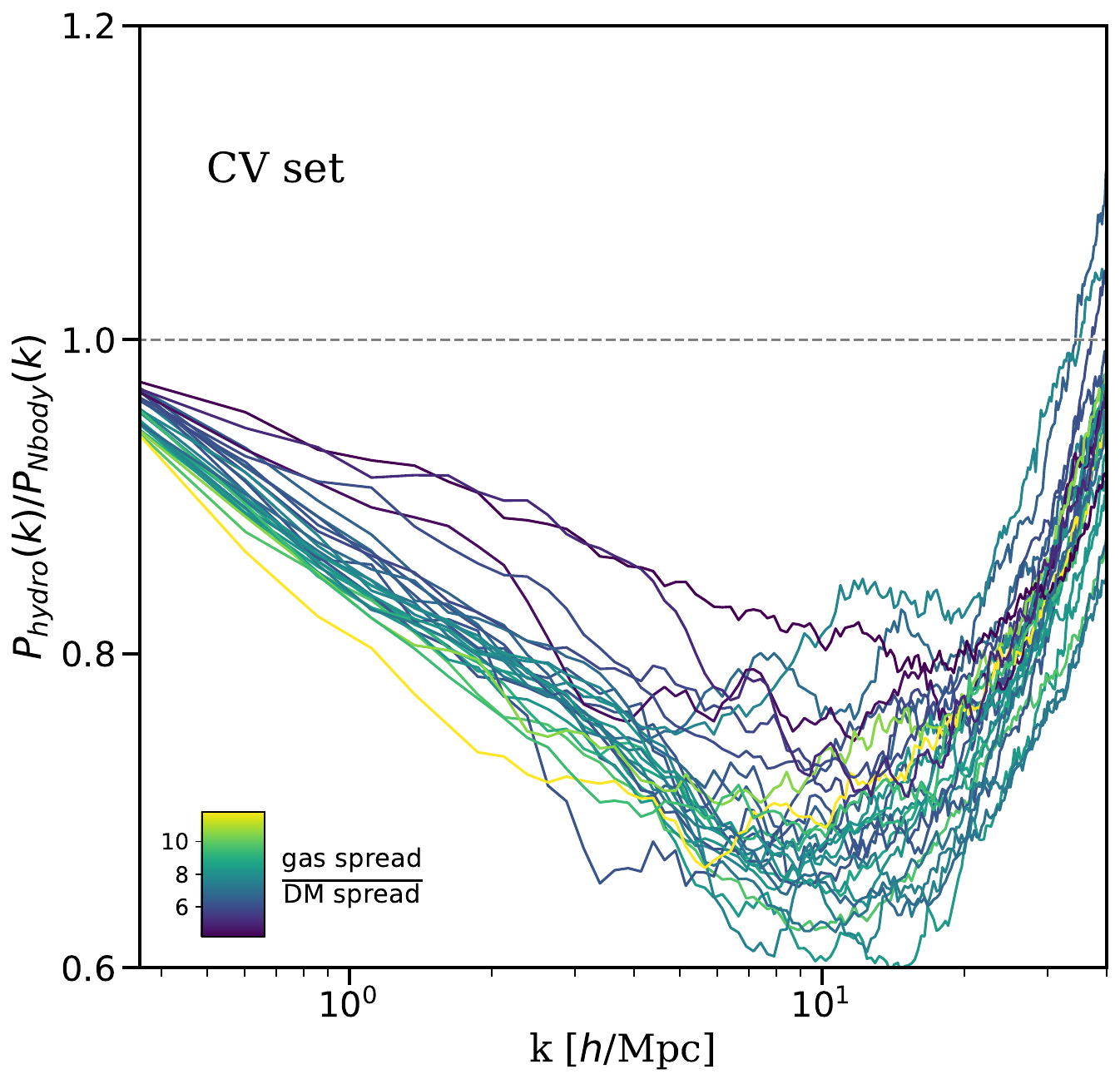}
    \caption{Same as Figure \ref{fig:1P_PS} but for the 27 simulation pairs of the \textsc{SIMBA} CV set, using fiducial simulation parameters and varied initial conditions. The strong correlation between suppression of power and baryonic spread holds on large scales ($k \lesssim 2\,h$\,Mpc$^{-1}$) but is lost on smaller scales due to cosmic variance.}
    \label{fig:CV_PS}
\end{figure}

\begin{figure}
	\includegraphics[width=\columnwidth]{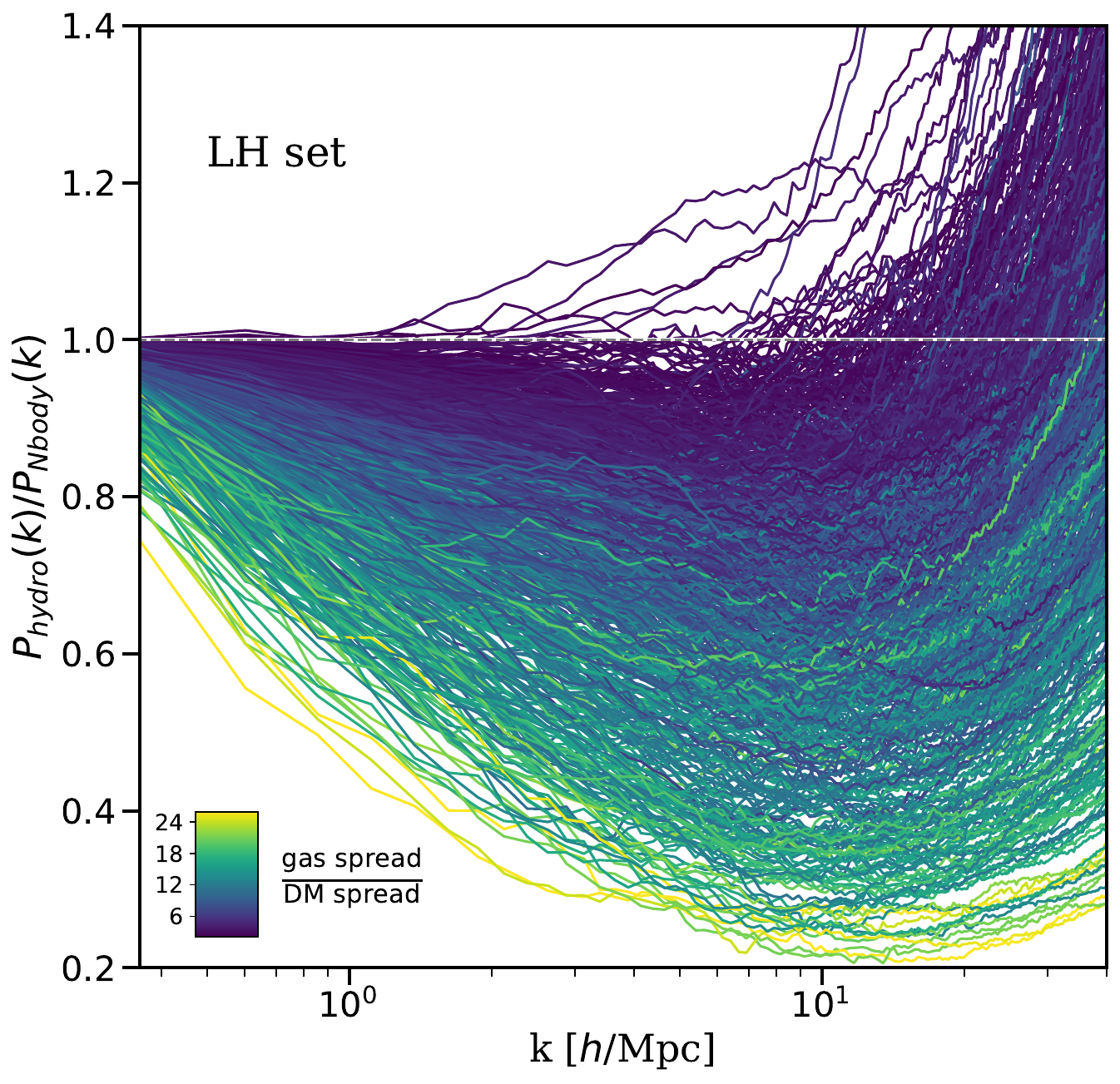}
    \caption{Same as Figure \ref{fig:1P_PS} but for the 1000 simulation pairs in the \textsc{SIMBA} LH set, implementing different cosmological and feedback parameters along with varying initial conditions. Despite the large scatter, gas spread is a good predictor of the suppression of matter clustering due to baryonic physics.}
    \label{fig:LH_PS}
\end{figure}

We next explore this relationship as a function of cosmic variance. Figure~\ref{fig:CV_PS} is similar to Figure~\ref{fig:1P_PS} but now showing $P_{\rm{hydro}}/P_{\rm{Nbody}}$ as a function of $k$ for the \textsc{SIMBA} CV set. Given the small $(\text{25\,Mpc}/h)^{3}$ simulated volumes in CAMELS, cosmic variance alone represents a significant amount of scatter in the amount of power suppression even for fiducial feedback parameters \citep{Villaescusa_Navarro_2021, Delgado_2023_powerspec, Ni_2023_CAMELSastrid}. Interestingly, the median gas spread remains a good predictor of the impact of feedback on the total matter power spectrum, capturing most of the variation of $P_{\rm{hydro}}/P_{\rm{Nbody}}$ due to cosmic variance on scales $k \lesssim 2\,h$\,Mpc$^{-1}$. Notably, this result holds even without normalizing the spread.
Figure \ref{fig:LH_PS} repeats this exercise for the full \textsc{SIMBA} LH set, where each of the 1000 simulations has different cosmological and astrophysical parameters in addition to different initial conditions.
In this case, there is significantly more scatter, but the relationship between total matter power spectrum suppression and the amount of baryon spreading continues to hold over a range of scales. 

\begin{figure}
	\includegraphics[width=\columnwidth]{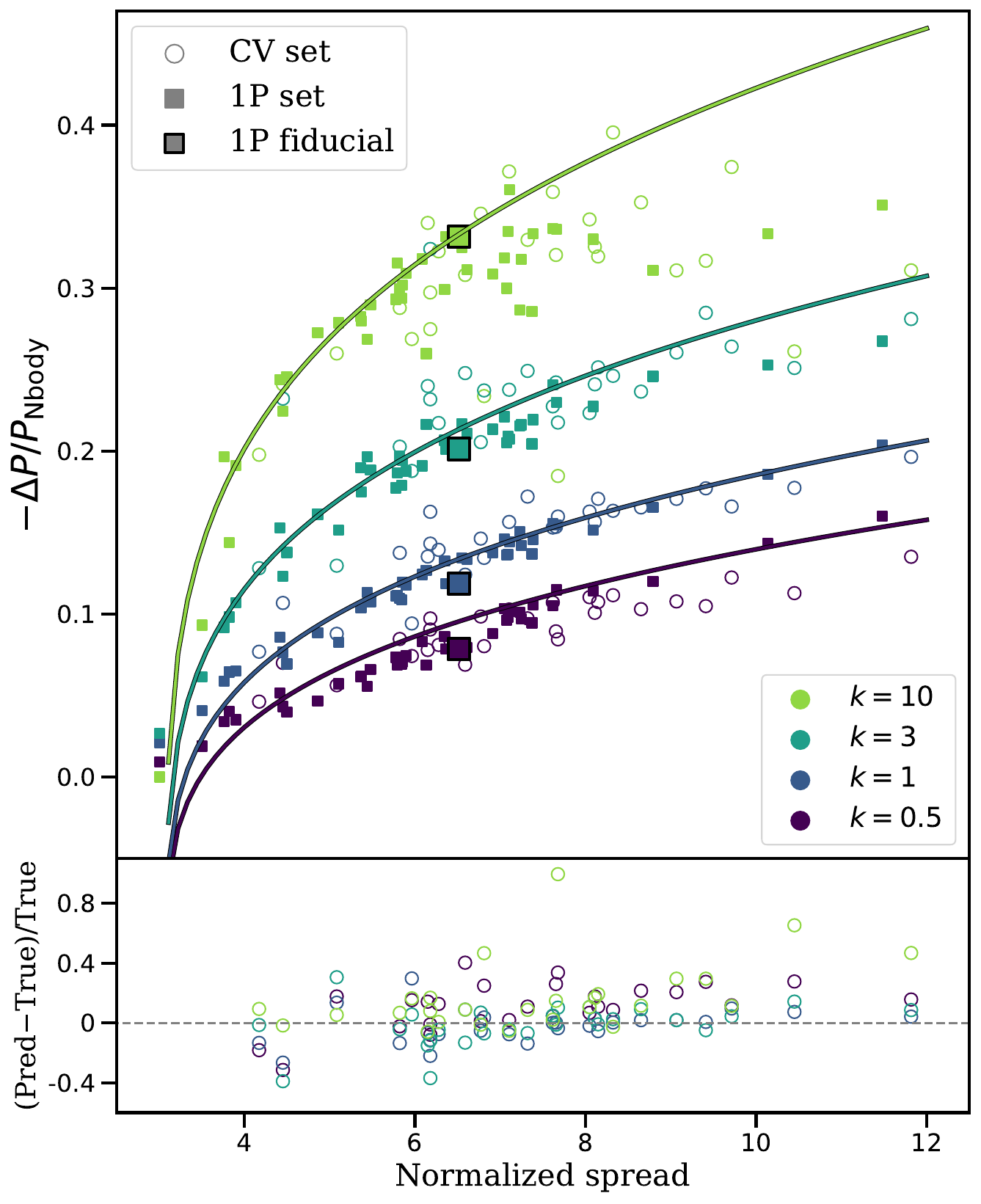}
    \caption{Fractional difference in power spectrum at $z=0$ as a function of the normalized spread (as described in Figure \ref{fig:1P_PS}) at different values of $k$ for both the six parameter 1P set (only feedback variations, squares) and CV set (circles) in \textsc{SIMBA} (top panel). The simulation with fiducial parameters from the 1P set is plotted with a larger, outlined square for each $k$ value. Best fit lines found via symbolic regression for the 1P simulations are plotted at each of these $k$ values. While only trained on the six parameter \textsc{SIMBA} 1P set simulations that varied feedback parameters, symbolic regression roughly captures the trends seen in the CV set as well. Errors for CV set predictions are shown in the bottom panel.}
    \label{fig:scatterCV1P}
\end{figure}

To examine this trend in further depth, we plot in Figure \ref{fig:scatterCV1P} the fractional difference in the power spectra against the normalized spread for different values of $k$, where we show $\Delta P/P \equiv (P_{\rm{hydro}} - P_{\rm{Nbody}}) / P_{\rm{Nbody}}$ for both the six parameter \textsc{SIMBA} 1P set (only feedback variations, squares) and CV set (circles) at four different values of $k$. The relationship between $\Delta P/P$ and baryonic spread is nearly linear at low values of $k$, while at higher $k$ values it takes on a more complicated form. Additionally, the data points from the CV set tend to have larger scatter, which increases with $k$ as expected from Figure~\ref{fig:CV_PS}. We use PySR symbolic regression \citep{pysr} with the six parameter \textsc{SIMBA} 1P set simulations that vary feedback parameters and find the following functional form to model the fractional difference in the total matter power spectrum due to baryonic effects as a function of both $k$ and the amount of spread:  
\begin{equation}
    -\Delta P/P = a_{1} \times (k \times (a_{2} \times S - a_{3}))^{1/4} - a_{4},
\end{equation}
where $S$ is the median normalized gas spread metric and $a_{1} = 0.25$, $a_{2} = 0.35$, $a_{3} = 1.09$, and $a_{4} = 0.14$ are the best-fit coefficients for the preferred functional form found by PySR. The errors from CV set predictions are shown in the bottom panel. Encouragingly, this simple expression roughly captures the dependence of $\Delta P/P$ on baryonic spread as a function of $k$ despite cosmic variance effects.

\subsection{Comparison to other CAMELS suites}
\label{Comparisons}

\begin{figure*}
	\includegraphics[width=\textwidth]{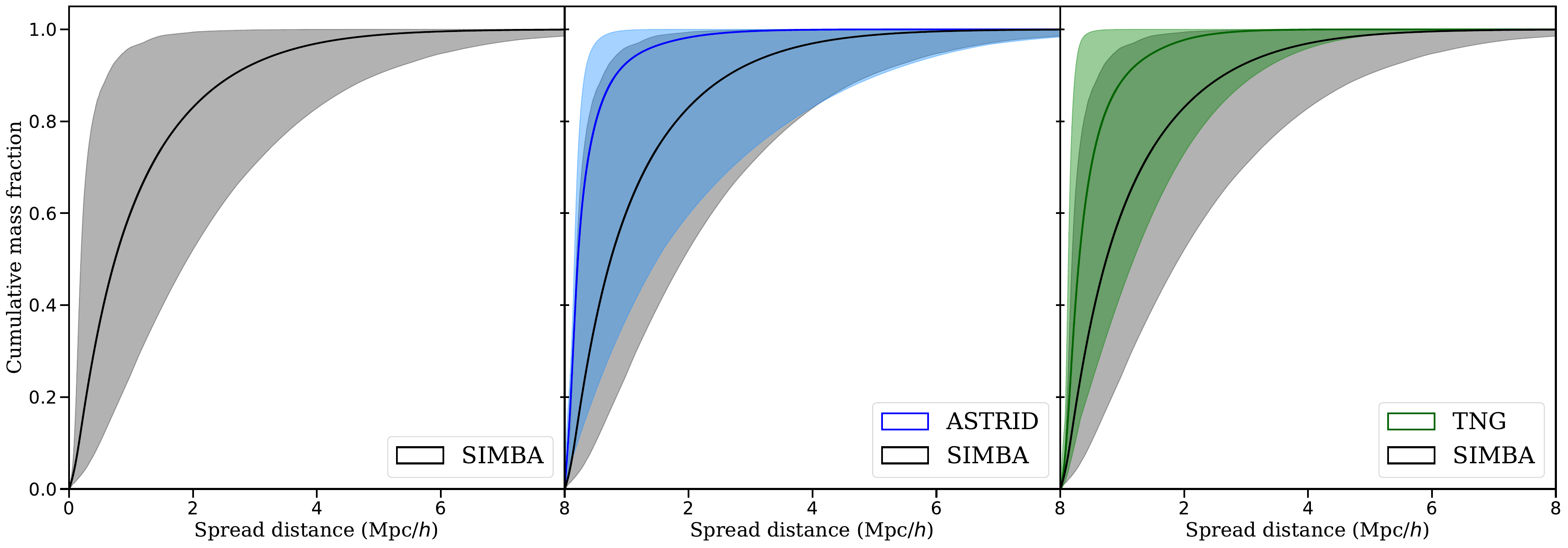}
    \caption{Cumulative mass distribution as a function of spread distance at $z=0$ for gas particles in the fiducial 1P simulation (solid lines) and the full LH set (SB28 for IllustrisTNG; shaded region) for \textsc{SIMBA} (grey; all panels), \textsc{ASTRID} (blue; middle panel), and IllustrisTNG (green; right panel). The fiducial \textsc{SIMBA} model spreads gas significantly further than \textsc{ASTRID} and IllustrisTNG, although the full range of gas spreading is comparable in the LH sets of \textsc{SIMBA} and \textsc{ASTRID}.}
    \label{fig:cumulativespreadfinal}
\end{figure*}

Here, we compare gas spreading between \textsc{SIMBA}, IllustrisTNG, and \textsc{ASTRID}. Figure \ref{fig:cumulativespreadfinal} shows the cumulative mass distribution of gas as a function of spread distance for each fiducial model (solid lines) and for the full range of parameter variations as given by either the LH set (for \textsc{SIMBA} and \textsc{ASTRID}) or the SB28 set (for IllustrisTNG). Clearly, the fiducial \textsc{SIMBA} simulation spreads gas the furthest ($\sim$40\% of gas in the simulated volume spreads further than 1\,Mpc/$h$ from its neighbouring dark matter), and the variations in the LH set predict a wide range of gas spread (anywhere from 1\% to 75\% of gas spreading further than 1\,Mpc/$h$). The fiducial \textsc{ASTRID} simulation appears to generally spread gas the least ($\sim$7\% spreading further than 1\,Mpc/$h$), but also predicts a wide variation in the LH set (0.2\% to 63\% for $>1$\,Mpc/$h$). The gas spread in the fiducial IllustrisTNG simulation is on a similar level to \textsc{ASTRID} ($\sim$11\% spreading further than 1\,Mpc/$h$). The range given by the SB28 set shows the smallest minimum and maximum spread (0.01\% to 57\%), but spans the largest logarithmic range. The increased gas spread seen in SIMBA relative to IllustrisTNG is in qualitative agreement with results found in \cite{Ayromlou_2022_closureradius}, where the ``closure radius'' (defined as the characteristic radius from a halo within which the enclosed baryon fraction equals the cosmic baryon fraction) is significantly larger in SIMBA relative to IllustrisTNG and EAGLE, indicating a greater redistribution of baryons.

\begin{figure}
	\includegraphics[width=\columnwidth]{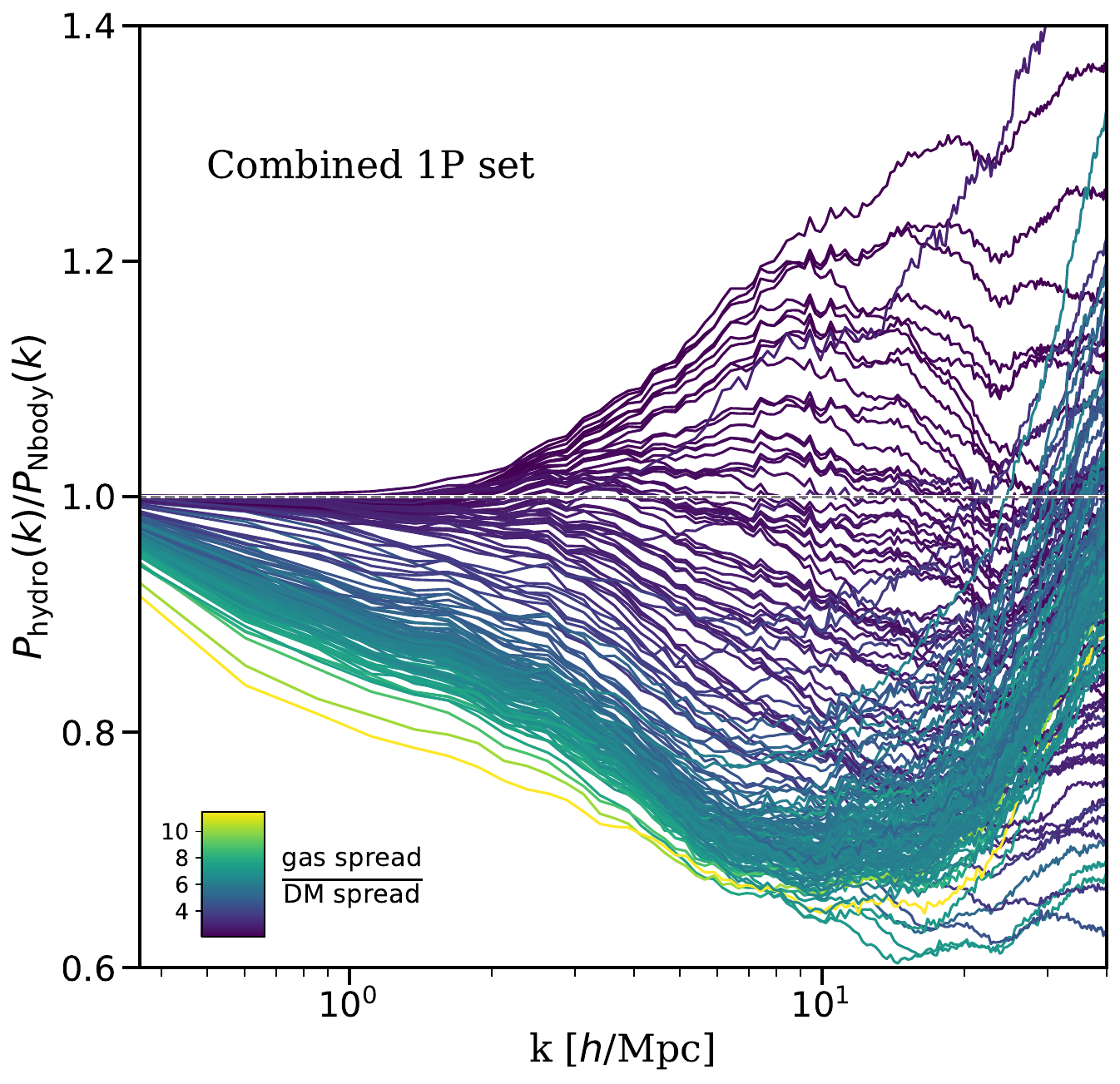}
    \caption{Same as Figure \ref{fig:1P_PS}, but now including simulations from the \textsc{ASTRID} 1P set, and 28 parameter 1P sets for IllustrisTNG and \textsc{SIMBA} that do not vary cosmological parameters. The correlation between suppression of power and gas spread is quite clear even across different galaxy formation models and variations of up to 23 astrophysical parameters.}
    \label{fig:full_1P_PS}
\end{figure}

\begin{figure}
	\includegraphics[width=\columnwidth]{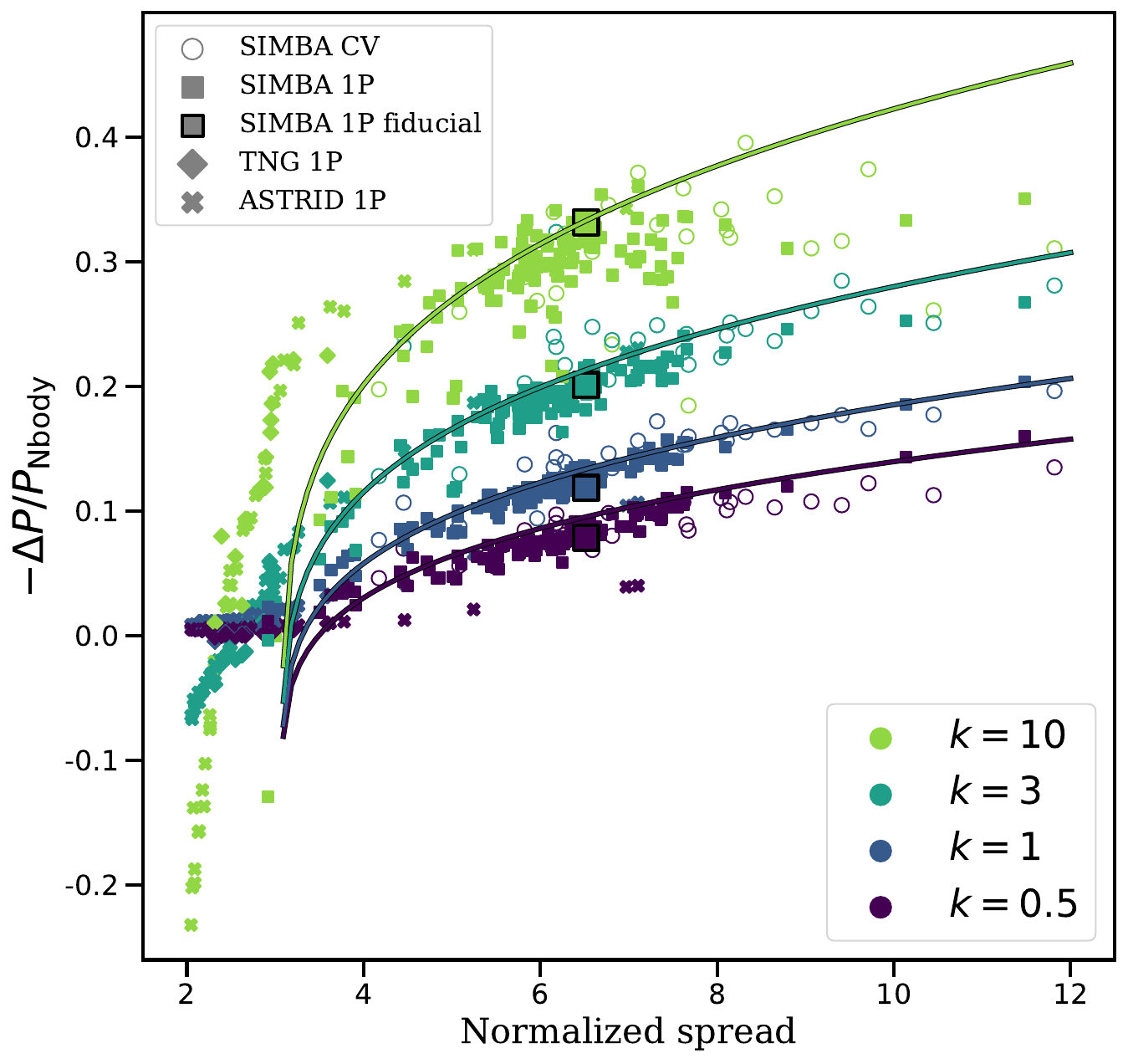}
    \caption{Same as Figure \ref{fig:scatterCV1P}, but now including simulations from the \textsc{ASTRID} 1P set, and 28 parameter 1P sets for IllustrisTNG and \textsc{SIMBA} that do not vary cosmological parameters. The same analytic function from symbolic regression is plotted (trained on six parameter \textsc{SIMBA} 1P simulations), which appears to be successful in reproducing the full \textsc{SIMBA} 1P results but does not capture the trends seen for IllustrisTNG and \textsc{ASTRID} at low spread values.}
    \label{fig:full_1P_scatter}
\end{figure}

We extend our investigation of the relationship between baryon spread and large-scale suppression of the matter power spectrum now as a function of galaxy formation model. Figure~\ref{fig:full_1P_PS} is an extension of Figure \ref{fig:1P_PS} that now includes simulations from the \textsc{ASTRID} 1P set and the full 28 parameter 1P sets for IllustrisTNG and \textsc{SIMBA} that do no vary cosmological parameters (180 simulations in total). Remarkably, we find a very clear correlation between the suppression of power on scales $k \lesssim 10\,h$\,Mpc$^{-1}$ and the gas spread metric regardless of galaxy formation model and variations of up to 23 astrophysical parameters. Once again, we investigate this trend quantitatively in Figure~\ref{fig:full_1P_scatter}, where we show $\Delta P/P$ as a function of gas spread for different values of $k$, as in Figure~\ref{fig:scatterCV1P} but now including also the \textsc{ASTRID} 1P set and the full 28 parameter 1P sets for IllustrisTNG and \textsc{SIMBA}. Here we reproduce the analytic function found with symbolic regression using only the six parameter \textsc{SIMBA} 1P simulations, which also seems to succeed in fitting the broader parameter variations in the full \textsc{SIMBA} 1P set. However, it is clear that this function does not quite match the \textsc{ASTRID} and IllustrisTNG simulations at all $k$, nor at small spreads. At small $k$, \textsc{ASTRID} suppresses power notably less than \textsc{SIMBA} simulations with comparable gas spread, and at large $k$, both \textsc{ASTRID} and IllustrisTNG show far more suppression than \textsc{SIMBA} when gas is not spreading far. Many \textsc{ASTRID} simulations spread gas by small amounts and in fact show an increase in power relative to $N$-body as the gas spread declines. There indeed appears to be a relationship between baryonic impact on the matter power spectrum and the spreading of baryons, but this trend can significantly depend on the galaxy formation model.

\section{Discussion}
\label{Discussion}
The decoupling of baryons from dark matter on cosmological scales represents a key signature of astrophysical feedback processes. As gas does not simply follow the gravitational pull of dark matter, $N$-body simulations do not tell the whole story. Systematic comparisons between $N$-body and hydrodynamic simulations allow for a controlled analysis of the role that feedback plays in the distribution of matter. It was shown in \cite{Borrow_2020} that baryonic matter can spread great distances away from the initial neighboring dark matter distribution in the \textsc{SIMBA} simulation. Here, we have used CAMELS to extend this analysis to a wide range of variations in cosmological and subgrid parameters in different plausible galaxy formation models, with the goal of encompassing the actual baryonic spread in the real Universe. Our results highlight the extent to which baryonic matter can cycle in and out of galaxies, be ejected from the circumgalactic medium (CGM), or even transferred to other halos \citep{Dave_2012_baryoncycle, Christensen_2016_baryoncycle, DAA_2017, Hafen_2019_FIREcgm, Wright_2020_EAGLE_baryoncycle, Mitchell_2020_EAGLE_baryoncycle, Hafen_2020_FIREcgm, Ayromlou_2022_closureradius}.

In the fiducial \textsc{SIMBA} model, 40\% of the gas mass in the entire simulated cosmological volume spreads further than 1\,Mpc, which is beyond the virial radius of all haloes in the simulation. Increasing AGN jet speed (the $A_{\rm{AGN2}}$ parameter in CAMELS) from lowest to highest increases this percentage from 25\% up to 55\%. This large-scale spreading of baryons may represent an important limitation for halo models and baryonification methods \citep{Seljak_2000_halo_model, Sembolini_2013_halo_model, Mead_2015_halo_model, Schneider_2015_baryonification, Schneider_2019_baryonification, Weiss_2019_baryonification} which consider only a redistribution of matter relative to $N$-body simulations within the scale of individual halos. Even with the weakest AGN feedback in \textsc{SIMBA}, more than 25\% of gas from halo Lagrangian regions (which should otherwise accrete onto haloes) end up spreading far out of haloes, which represents a significant source of uncertainty for baryonification models.

We have shown that dark matter itself spreads (relative to its initial neighbouring dark matter distribution) by the largest amount within and around massive haloes, as they are the largest sources of gravitational potential. Dark matter spreading increases when haloes get larger, as chaotic dynamics can make initial particle neighbor trajectories diverge on scales comparable to the splashback radius \citep{Diemer_2014_splashback, Adhikari_2014_splashback, More_2015_splashback, Mansfield_2017_splashback}. We see larger haloes forming when increasing the values of $\Omega_{\rm{m}}$ and $\sigma_{8}$ (as expected, see\citealt{Villaescusa_Navarro_2021}), but otherwise large-scale dark matter spreading seems to be roughly independent of these cosmological parameter variations. 
One exception is the slight decrease in dark matter spread in haloes of equal mass as $\Omega_{\rm{m}}$ (and $\sigma_8$ for lower-mass haloes) increases (Figure \ref{fig:dm_halo_masses}), which may be explained by variations in halo concentration. Higher values of $\Omega_{\rm{m}}$ and $\sigma_8$ increase halo concentration at fixed virial mass \citep{Dooley_2014_haloconcentration}, which would result in haloes with more mass concentrated in the central region and thus reduced amount of dark matter spreading within them. 
Gas largely follows the same trend as $\Omega_{\rm{m}}$ and $\sigma_{8}$ increase but experience additional interactions. Generally, larger haloes may increase spread by generating stronger feedback (due to the presence of more massive central black holes) and having chaotic particle trajectories over larger scales, but also host baryons that collapse to higher densities in the central galaxy and end up spreading very little.

The dependence of gas spread on feedback parameters shows complex non-linear effects. A clear result in Figure \ref{fig:gas_spread} is that increasing the strength of AGN feedback in \textsc{SIMBA} (momentum flux and jet speed) significantly increases gas spread, while increasing the stellar feedback efficiency either yields mixed results (mass loading factor) or a significant decrease in gas spread (wind speed). One possible explanation for this non-intuitive result is that there is significant non-linear interaction between stellar and AGN feedback, as seen in previous works \citep{Booth_schaye_2013_SN_AGN, van_Daalen_2020, Nicola_2022_sumstat, Delgado_2023_powerspec}. Analyses of high resolution FIRE zoom-in simulations show that strong stellar feedback can limit early black hole growth by continually ejecting material away from the nuclear region (\citealt{DAA_2017_BHgrowth,Catmabacak_2022_FIREBH, Byrne_2023_FIREBH}; see also \citealt{Dubois_2015_SethBH, Bower_2017_EAGLEBH, Habouzit_2017_ramsesBH, Lapiner_2021_NewHorizonsBH}), which can therefore reduce the impact of AGN feedback. Indeed, \cite{Ni_2023_CAMELSastrid} showed that increasing the SNe wind speed parameter ($A_{\rm{SN2}}$) in SIMBA greatly decreased the quantity of massive black holes. \cite{Borrow_2020} found that gas particles tagged as having directly interacted with AGN jets in \textsc{SIMBA} were spread significantly further than particles that only directly interacted with stellar feedback. Furthermore, by contrasting with a ``No-Jet'' \textsc{SIMBA} simulation, it was shown that gas does not spread to large distances without AGN jets and instead spreads on the same level as dark matter. Given these effects, our results suggest that increasing the speed of galactic winds  suppresses the overall output of AGN jets and therefore their ability to redistribute gas over large scales. 

Massive dark matter haloes are responsible for both very small and extremely large baryonic spreads owing to the competing effects of radiative cooling (allowing gas to collapse down to halo centers) and strong AGN feedback (ejecting gas to large scales). This dichotomy in the fate of gas can be clearly seen for gas particles that belong to halo Lagrangian regions at the initial conditions (Figure \ref{fig:gas_spread_haloes}). Gas particles that remain inside of their parent halos at $z=0$ spread very little, with minimal dependence on feedback parameters, while Lagrangian region gas outside of parent halos at $z=0$ show significantly larger and feedback-dependent spreads. Investigating the large-scale spatial distribution of gas as a function of the amount of spread provides further support for this picture (Figure \ref{fig:gas_AGN2_25}), where the least spread gas is constrained to halo centers while the most spread gas is spread out around large haloes and filaments \citep[in agreement with][]{Borrow_2020}. As the jet speed increases, the least spread gas is even more tightly constrained to halo centers, and the most spread gas is even more diffusely spread out over a large fraction of the simulated volume in CAMELS. This depiction is in agreement with our finding in Figure~\ref{fig:gas_spread_haloes} that the in-halo gas spread variation (25th-75th percentile shaded region) decreases while the out-of-halo gas spread increases with higher AGN jet speed.

We have shown that the amount of baryonic spread in simulations is closely related to the overall impact of feedback on the total matter power spectrum. Previous authors have investigated the impact of baryons on the matter power spectrum and found that it can be significantly ``contaminated'' by non-linear baryonic effects relative to dark matter-only simulations \citep{van_Daalen_2011, Chisari_2019, van_Daalen_2020, Delgado_2023_powerspec, Pandey_2023_powerspec}. This contamination typically results in a reduction of power on large scales in hydrodynamic simulations as stellar feedback and, in particular, AGN feedback redistribute gas far from haloes they would otherwise reside in or around. 
With a large extended set of $\sim$200 simulations with identical initial conditions and varying up to 23 astrophysical parameters in the \textsc{SIMBA}, IllustrisTNG, and \textsc{ASTRID} models \citep{Ni_2023_CAMELSastrid}, we have shown that there is a tight correlation between the suppression of power on scales $k \lesssim 10\,h$\,Mpc$^{-1}$ and the large-scale baryon spread (Figure \ref{fig:full_1P_PS}). Generally, simulations that spread baryons further relative to their initial neighboring dark matter distribution show a greater suppression of power on large scales, regardless of the specific galaxy formation model and feedback parameter variations.

Due to the small simulated volumes in CAMELS, cosmic variance can have significant effects on many measured quantities. While all of the \textsc{SIMBA} CV set simulations implement identical feedback parameters, different initial conditions may result in a different population of haloes for which the same feedback model can have widely different effects. In particular, our results highlight the extent to which cosmic variance in a (25\,Mpc/$h)^3$ volume can play a role in the large-scale spreading of baryons, with some \textsc{SIMBA} CV simulations showing a median gas spread twice that of other simulations with identical parameters.
Previous works in CAMELS have partially mitigated the limitation of small simulated volumes by finding good predictors of cosmic variance. \cite{Nicola_2022_sumstat} reduced the effect of cosmic variance on neural networks trained to constrain cosmological and astrophysical parameters from electron density power spectra by incorporating a parameter encoding the distribution of halo masses for each input simulation. 
\cite{Thiele_2022_paramconstraint} quantified the constraining power of spectral distortion measurements for baryonic feedback models and reduced sample variance in CAMELS by deriving a correction factor based on scaling the measured halo mass functions in CAMELS to that of reference large volume simulations. 
\cite{Delgado_2023_powerspec} trained a random forest regressor to predict the impact of baryonic effects on the matter power spectrum given halo baryon fractions and increased its predictive power significantly by including a form of the halo mass function as input feature, partially mitigating cosmic variance effects.
Interestingly, the normalized gas spread metric is an excellent predictor of the effects of cosmic variance on the inferred impact of feedback on the matter power spectrum, where the amount of power suppression in \textsc{SIMBA} CV simulations with identical feedback parameters is tightly correlated with baryon spreading up to scales $k \lesssim 2\,h$\,Mpc$^{-1}$ (Figure \ref{fig:CV_PS}). 

The way feedback processes are implemented, and thus the impact they have on the distribution of matter, can vary greatly between hydrodynamic simulation models. \cite{Chisari_2019} compared the impact of baryons on the matter power spectrum at $z=0$ in the fiducial models of a handful of cosmological hydrodynamic simulations, including Horizon-AGN \citep{Dubois_2014_HorizonAGN}, MassiveBlack-II \citep{Khandai_2015_massiveblack}, OWLS \citep{Schaye_2010_owls}, cosmo-OWLS \citep{LeBrun_2014_cosmoOWLs}, EAGLE \citep{Schaye_2015_EAGLE}, BAHAMAS \citep{McCarthy_2017_bahamas, McCarthy_2018_bahamas}, Illustris \citep{Vogelsberger_2014_illustris2, Vogelsberger_2014_illustris, Genel_2014_illustris}, and IllustrisTNG \citep{Springel_TNG_Pk_2018}. They found that the suppression of power relative to $N$-body simulations can range from 10--30\% in the different models at wave numbers from a few up to $20\,h\,\rm{Mpc^{-1}}$. More recently, \cite{van_Daalen_2020} found similar results when quantifying the impact of baryonic physics on matter clustering for nearly 100 simulations from the OWLS, cosmo-OWLS, and BAHAMAS models that varied cosmological and feedback parameters. 
Our results further emphasize that the distribution of matter in cosmological hydrodynamic simulations strongly depends on both the simulation model and the strength of feedback. Figure \ref{fig:cumulativespreadfinal} highlights this fact, with the spreading of baryons varying widely throughout thousands of CAMELS simulations with different parameters, initial conditions, and galaxy formation models. This large library of simulations enables the development of machine learning algorithms that can quantify baryonic uncertainties and marginalize over them for cosmological parameter inference \citep{Villaescusa_Navarro_2021, Perez_2022_camelsSAM, Ni_2023_CAMELSastrid}, as well as devise observational probes that can help constrain baryonic physics. For example, \cite{Nicola_2022_sumstat} used CAMELS to investigate the electron density power spectrum (measurable through kinematic Sunyaev-Zel’dovich observations or Fast Radio Burst dispersion measures) as a means to break the baryon-cosmology degeneracy \citep[see also][]{YongseokJo_2023_camels} and improve theoretical models of the impact of baryonic feedback on the matter power spectrum, and \cite{Pandey_2023_powerspec} used Dark Energy Survey weak lensing and Atacama Cosmology Telescope thermal Sunyaez-Zel'dovich effect measurements together with models trained on CAMELS to constrain the impact of feedback on matter clustering. Constraints such as these will be extremely valuable for extracting the maximum information out of upcoming cosmological surveys to further constrain the fundamental parameters of the Universe.

Previous authors have also noted a connection between lower halo baryon fraction $f_{\rm b}$ and increased baryonic impact on the matter power spectrum \citep{van_Daalen_2020, Nicola_2022_sumstat, Pandey_2023_powerspec, Delgado_2023_powerspec}. Our results connecting baryonic spread and suppression of power across a large library of models are in agreement with these findings. Simulations forming halo populations with lower $f_{\rm b}$ require more baryons to be ejected out of haloes, which results in greater baryonic spread. The baryon fraction of massive haloes ($M_{\rm halo} \sim 10^{14}\,\rm{M_{\odot}}$) is particularly well correlated with the suppression of the matter power spectrum, and \cite{van_Daalen_2020} formulated an empirical model to predict power suppression as a function of $f_{\rm b}$ which is accurate up to $k \lesssim 1\,h$\,Mpc$^{-1}$ across a number of galaxy formation models. However, \cite{Delgado_2023_powerspec} showed that this model cannot capture the $f_{\rm b}$--power suppression relation in \textsc{SIMBA}, expanding this work to thousands of CAMELS simulations across the full halo mass range ($10^{10}\,\rm{M_{\odot}} \le M_{halo} \le 10^{14}\,\rm{M_{\odot}}$) to train a random forest regressor capable of predicting the baryonic impact on the matter power spectrum well into the non-linear regime.
We have shown that a simple analytic function found by symbolic regression can roughly capture the suppression of the matter power spectrum as a function of baryon spread also into the non-linear regime for \textsc{SIMBA} simulations varying 23 astrophysical parameters despite training on simulations varying only 4 feedback parameters (Figure~\ref{fig:scatterCV1P}). However, the same model cannot accurately capture this relationship for IllustrisTNG and \textsc{ASTRID}, particularly in simulations with lower baryon spread compared to that represented in the \textsc{SIMBA} simulations (Figure~\ref{fig:full_1P_scatter}). 
This emphasizes the need to construct robust models relative to changes in the galaxy formation implementation, which is a common difficulty of many current models \citep[for further discussion and examples of robust models, see][]{Villaescusa-Navarro_2021_marginalization_fields2, Villaescusa-Navarro_2021_marginalization_fields,Villaescusa-Navarro_2022_OneGalaxy,deSanti_2023_fieldlevelgalaxies, Echeverri_2023_camels, Shao_2023a, Shao_2023b}.

\section{Conclusions}
\label{Conclusions}
We have explored in detail the use of a novel {\it cosmological spread metric} to describe the 
 redistribution of dark and baryonic matter on large scales owing to gravitational dynamics and feedback from astrophysical sources, providing a unifying framework to interpret and quantify the impact of baryonic effects on the total matter power spectrum regardless of the specific feedback prescriptions used in different galaxy formation models. Our main results can be summarized as follows:

\begin{itemize}
  \item Dark matter spreads relative to the initial neighboring matter distribution owing to chaotic gravitational dynamics, with the largest spread distances occurring in and around massive halos.  As expected, dark matter spreading increases with $\Omega_{\rm{m}}$ and $\sigma_{8}$ following the formation of higher mass haloes in simulations.
  
  \item On average, gas spreads much further than dark matter due to astrophysical feedback effects. Radiative cooling can allow gas to lose energy and fall to lower bound orbits at the centers of haloes, but gas impacted by feedback can be ejected to large distances. This dichotomy of gas cooling and feedback yields a large variation in spread distances for gas inside of halo Lagrangian regions at the initial conditions.
  
  \item Increasing the efficiency of AGN feedback increases the spread of gas, but increasing the stellar feedback efficiency can decrease the spread of gas. This supports the notion that AGN feedback is the dominant component spreading baryons to large distances while stronger stellar feedback may inhibit black hole growth and therefore reduce the impact of AGN feedback.
  
  \item The baryonic spread metric is a good predictor of the global impact of feedback on the large scale distribution of matter as described by the ratio of the matter power spectrum in hydrodynamic and $N$-body simulations, with larger baryonic spread driving stronger suppression of power on large scales. 
  
  \item Using symbolic regression, we have found a simple analytic function that captures the matter power suppression as a function of wave number and baryonic spread in simulations varying $>$20 astrophysical parameters in the \textsc{SIMBA} model, while extrapolating to IllustrisTNG and \textsc{ASTRID} simulation variations that spread baryons significantly less than \textsc{SIMBA} remains a challenge.
\end{itemize}

The extent to which matter is redistributed by feedback processes is significant and can contribute to uncertainties in approximation methods that do not model baryonic physics explicitly, while predictions from cosmological hydrodynamic simulations can vary widely depending on the choice of feedback parameters and model implementation.  Besides providing a clear physical interpretation for the impact of baryonic physics on the matter power spectrum, the simplicity of the spread metric makes it a useful summary statistic to characterize the global efficiency of feedback in galaxy formation simulations.

\section*{Acknowledgements}
We thank Dylan Nelson and Joop Schaye for useful suggestions that helped improve the paper. 
The CAMELS simulations were performed on the supercomputing facilities of the Flatiron Institute, which is supported by the Simons Foundation. DAA acknowledges support by NSF grants AST-2009687 and AST-2108944, CXO grant TM2-23006X, Simons Foundation Award CCA-1018464, and Cottrell Scholar Award CS-CSA-2023-028 by the Research Corporation for Science Advancement. DN acknowledges support from the NSF AST-2206055 grant.

\section*{Data Availability}
The CAMELS simulations are publicly available (see \citealt{CAMELS_data_release} and \citealt{Ni_2023_CAMELSastrid}). More information and instructions for downloading the data can be found at https://camels.readthedocs.io.


\bibliographystyle{mnras}
\bibliography{baryonspread}



\appendix
\section{Dark matter spread in selected halo mass ranges}
\label{appendix}
\begin{figure}
	\includegraphics[width=\columnwidth]{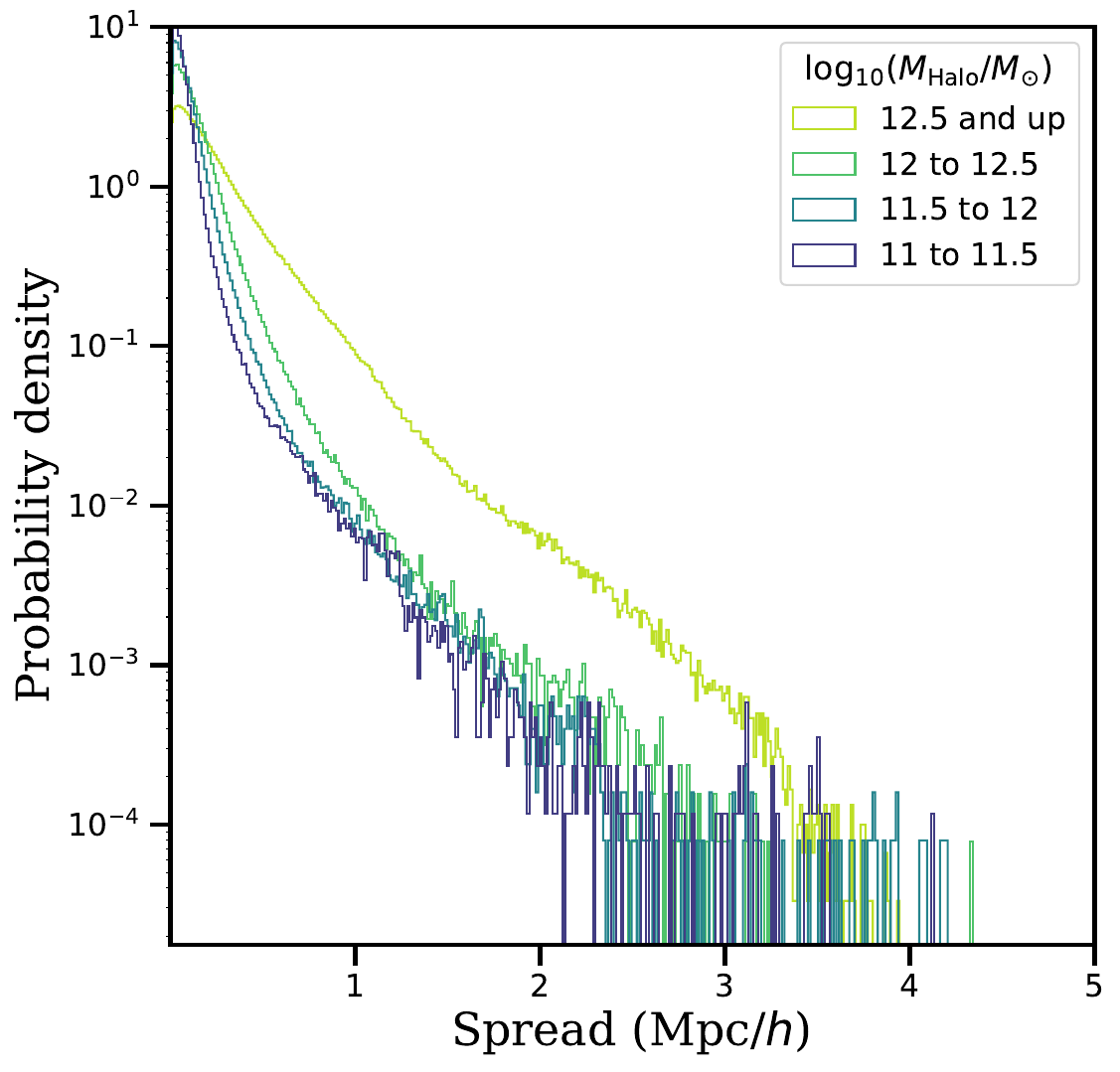}
    \caption{Distribution of spread distances for dark matter particles in haloes of selected mass ranges in the fiducial $N$-body simulation at $z=0$. Larger haloes generally spread dark matter farther, but cases of extremely large spreads appear to be independent of halo mass.}
    \label{fig:dm_halo_spread}
\end{figure}

The spreading of dark matter particles within haloes shows a strong dependence on halo mass, but seemingly little direct dependence on cosmological parameters (Figure \ref{fig:dm_halo_masses}). In Figure \ref{fig:dm_halo_spread}, we show the full spread distributions of dark matter particles in haloes of the same mass ranges used in Figure \ref{fig:dm_halo_masses} for the fiducial $N$-body 1P simulation. As expected, dark matter particles in more massive haloes spread farther from their initial neighbors. The most extreme cases of spread can occur in any mass range, corresponding to particles whose initial neighbor ended up very far from the halo.

\bsp	
\label{lastpage}
\end{document}